\documentclass[hidelinks,12pt]{article}
\usepackage[utf8]{inputenc}
\usepackage[english]{babel}
\usepackage[a4paper,lmargin={1.7cm},rmargin={1.7cm},tmargin={2.15cm},bmargin={2.15cm}]{geometry}
\usepackage{color}
\usepackage{mathptmx}
\usepackage{subcaption}
\usepackage{amssymb}
\usepackage{amsmath,amsfonts,amsthm,bm}
\usepackage{graphicx, multicol}
\usepackage{epstopdf}
\usepackage[onehalfspacing]{setspace}
\usepackage{nth}
\usepackage[dvipsnames]{xcolor}
\usepackage{qtree}
\usepackage{verbatim}
\usepackage{csquotes}

\usepackage{pdflscape}
\usepackage{afterpage}

\usepackage{tabularx}
\usepackage{array}
\newcolumntype{b}{X}
\newcolumntype{s}{>{\hsize=.5\hsize}X}
\newcolumntype{Y}{>{\hsize=11.5\hsize}X}
\newcolumntype{G}{>{\centering\arraybackslash\hsize=4.25\hsize}X}

\usepackage{mathpazo} 
\usepackage[scaled]{helvet} 
\usepackage{courier} 
\normalfont
\usepackage[T1]{fontenc}
\usepackage{listings}
\usepackage{colortbl}
\definecolor{siena}{rgb}{0.91,0.45,0.32} 	

\usepackage[bookmarks,colorlinks]{hyperref}

\usepackage{cleveref}

\usepackage{multirow}

\usepackage{makecell}

\usepackage{empheq}

\usepackage[makeroom]{cancel}

\usepackage[export]{adjustbox}

\usepackage{colortbl}
\definecolor{siena}{rgb}{0.91,0.45,0.32} 	

\hypersetup{
  colorlinks,
  citecolor=PineGreen,
  linkcolor=red,
  urlcolor=blue}
  
\usepackage{float}
\usepackage{threeparttable}
\usepackage{booktabs}
\usepackage{titling}
\usepackage{natbib}
\usepackage{bibunits}

\thanksmarkseries{arabic}

\renewenvironment{abstract}
 {\small
  \begin{center}
  \bfseries \abstractname\vspace{-.5em}\vspace{0pt}
  \end{center}
  \list{}{
    \setlength{\leftmargin}{1.5cm}    \setlength{\rightmargin}{\leftmargin}  }  \item\relax}
 {\endlist}
 
\begin{document}
\vspace{-3cm}

\title{\LARGE {On Spurious Causality, CO$_2$, and Global Temperature}} 
\author{\hspace{-0.75cm} Philippe Goulet Coulombe\thanks{%
Department of Economics,  \href{mailto:gouletc@sas.upenn.edu}{{gouletc@sas.upenn.edu}}. We thanks an anonymous referee who suggested the use of information flows in an earlier paper of ours. We are grateful to Boyuan Zhang for many helpful comments and suggestions.} \qquad \qquad \hspace{0.15cm} Maximilian G\"obel}
\date{\vspace{-0.4cm} \hspace{0.95cm} University of Pennsylvania \qquad \hspace{-0.1cm} ISEG - Universidade de Lisboa  \\[2ex]%
\small
First Draft: October 10, 2020 \\
This Draft: \today \\ 
\vspace{0.4cm}
\large
 } 

\newgeometry{left=1.8cm, right = 1.8cm, bottom = 2.0cm, top = 2.0cm}
\maketitle\thispagestyle{empty}
\begin{abstract}

\noindent \textcolor{PineGreen}{Stips, Macias, Coughlan, Garcia-Gorriz,  and Liang (2016)} (\textit{Nature Scientific Reports}) use information flows \citep{Liang2008,Liang2014} to establish causality from various forcings to global temperature. We show that the formulas being used hinges on a simplifying assumption that is nearly always rejected by the data. We propose an adequate measure of information flow based on Vector Autoregressions, and find that most results in \cite{StipsEtAl2016} cannot be corroborated. Then, it is discussed which modeling choices (e.g., the choice of CO$_2$ series and  assumptions about simultaneous relationships) may help in extracting credible estimates of causal flows and the transient climate response simply by looking at the joint dynamics of two climatic time series.
\end{abstract}

\vspace{6.5cm}

\restoregeometry



\clearpage


\clearpage

\newpage
\pagenumbering{arabic}

\section{Introduction}
Causality is fundamental to science. While causal statements can reasonably be made without hardship in controlled environments, things are far less straightforward when only observational data is available. Answering the question of what happens to $Y$ if one intervenes on $X$ is compromised by the simple fact that $X$ and $Y$ were not generated by exogenously modulating $X$, but by (often) endogenous interactions between the two variables. 

Yet, for many scientific interrogations of capital importance -- like the relationship between greenhouse gases and global temperature, only observational data is available. How does different forcings cause global mean surface temperature anomalies (GMTA)? In a \href{https://www.nature.com/articles/srep21691/metrics}{popular} article, \textcolor{PineGreen}{Stips, Macias, Coughlan, Garcia-Gorriz,  and Liang (2016)} (henceforth, SMCGL) set up to answer the question using \emph{information flow} (IF) from one variable to another for bivariate stochastic dynamical systems -- a methodology developed in \cite{Liang2008,Liang2014,Liang2015,Liang2016}. Authors make grand claims about the technique being able to extract "true" and "rigorous" causality, which  clash with the usually somber tone of causal analysis. 

We show that the IF methodology will \textit{not} extract causality for the near-universe of bivariate stochastic dynamic systems estimated on real data -- with those of SMCGL included. In a nutshell, this occurs because \cite{Liang2008} formulas being used throughout, assume that a certain statistical quantity is 0 when it is not. This can be tested, and it is rejected almost all the time. More precisely, \cite{Liang2014} formula assumes that when conditioning on the past state of the system ($X_{t-1}$ and $Y_{t-1}$), the remaining variation in $X_{t}$ and $Y_{t}$ (the innovations driving the system) are \textit{uncorrelated}. In SMCGL, where the time unit $t$ is one year, this implies that forcings and GMTA are uncorrelated \textit{within the same year}, conditional on last year's values of both. That is unlikely to happen, and the data will testify to that. The assumption is presented as "linearity" \citep{Liang2014}, but it has little to do with it. Rather, it assumes that the system is identified, meaning that the joint dynamics of the time series data fall on the knife edge case where causality can indeed be claimed from the data without any external assumption \citep{sims1980}. The problem is that for discretely sampled time series, the assumption is almost always rejected by lower-frequency data.\footnote{In \cite{tawia2019time}, IF are used on daily data, which can alleviate the problem \textit{if there are no intra-day relationships}. This last condition is something that should be verified, not assumed.} In short, SMCGL assume there to be no identification problem (in the sense of \citealt{sims1980}), and find no identification problem.

In this note, we detail the consequences of this high-standing omission on IF measures and SMCGL's results. First, in simulations (using data generating processes proposed by \cite{Liang2014}) of which we know the true causality structure, it is shown that IF will often conclude that $X_t$ is largely causing $Y_t$, when in fact the reverse is true. Second, we reconsider a key part of SMCGL's study where the authors investigate the causal structure between different forcings and GMTA. Using an appropriate methodology that accounts for the correlated innovations, it is found that in most instances, the data by itself cannot back SMCGL's claims.  That is, unlike what the authors have put forward, it is not possible (within this framework) to claim that many forcings are causing GMTA's increase as a direct implication of the data. In other words, from the data \textit{alone}, it is impossible to discriminate between certain forcings mostly causing GMTA or the reverse. External assumptions based on physical knowledge could remedy that. Or different data. We show that in the case of CO$_2$, results in accord with the scientific consensus can be obtained when using concentration directly rather than its radiative content. 

The alternative methodology that we use are structural Vector Autoregressions (SVAR) which are simultaneous dynamic systems of equations. They can characterize a linear dynamic system in discrete time. The methodology was introduced to macroeconomics by \cite{sims1980} and is now so in many fields, ranging from neuroscience \citep{chen2011vector} to climate \citep{VARCTIC}. To document its reliability for this application, we report implied transient climate response (TCR) estimates of our alternative methodology. We find that those (i) largely depend  on the necessary assumptions made about the simultaneous (with one year) impact of CO$_2$ forcing on GMTA but (ii) are in the range of recent estimates \citep{OttoEtAl2013,MontamatStock2020} if one assumes simultaneous causality running from  CO$_2$ forcing to GMTA.


Section \ref{sec:theory} briefly review IFs and explain where the problematic assumption occurs. It also discusses relevant notions of Vector Autoregressions (VARs, \citealt{sims1980}) as a comprehensive framework to think about causality in multivariate time series systems. Section \ref{sec:simuls} display IFs possibly spurious behavior using simulated data where the true causality is known. Section \ref{sec:empirics} revisit the question of causality between different forcings and GMTA using appropriate tools. Section \ref{sec:con} concludes.

\section{Information Flows, Non-Innocuous Assumptions and VARs}\label{sec:theory}

In this section, we review the basics of IFs as applied in empirical work, pin down the problematic assumption, explain why it is harmful through the lenses of a VAR, and propose an alternative measure of IF based on the VAR. 

\cite{Liang2008} considers the  data generating process (DGP)
\begin{gather} \label{equ:DGP_IF}
d\textbf{X} = \textbf{F}\left(\textbf{X},t\right)dt + \textbf{B}\left(\textbf{x},t\right)d\textbf{W} \;,
\end{gather}

\noindent where $\textbf{X} = \left(X_1,X_2\right) \in \mathbb{R}^2$ are the state variables and $\textbf{F} = \left(F_1,F_2\right)$. $\textbf{W} = \left(W_1,W_2\right)$ is a standard Wiener process, with $\Delta \textbf{W} = \frac{d\textbf{W}}{dt}$ and $E\left(\Delta W_i\right) = 0$ and $E\left(\Delta W_i\right)^2 = \Delta t$. $\textbf{B}$ is a $2 \times 2$ matrix and its entries $b_{ij}$ govern how perturbations instantaneously impact the system. At this point, the only non-innocuous assumption is that of a bivariate system. The information flow from $X_2$ to $X_1$, $T_{2 \rightarrow 1}$, is then defined as
\begin{gather} \label{equ:IFequation}
T_{2 \rightarrow 1} = \frac{dH_1}{dt} - \frac{dH_{1\bcancel{2}}}{dt}
\end{gather}
\noindent  where $\frac{dH_1}{dt}$ is the evolution of the marginal entropy of $X_1$ and $\frac{dH_{1\bcancel{2}}}{dt}$ denotes $\frac{dH_1}{dt}$ where the spillovers from $X_2$ are excluded. After lengthy derivations and reasonable assumptions (like that $W$'s, soon to be called structural shocks, are uncorrelated) it is obtained that 
\begin{align} \label{equ:IFequation_general}
T_{j \rightarrow i} = -E\left[\frac{1}{\rho_i} \frac{\partial \left(F_i \rho_i\right)}{\partial X_i}\right] + \frac{1}{2} E\left[\frac{1}{\rho_i} \frac{\partial^2\left(g_{ii} \rho_i\right)}{\partial X^2_i}\right] \;,
\end{align}
\noindent where $\rho$ is the joint probability density of variables $X_i$ and $X_j$, and $\rho_i$ denotes the marginal density of series $X_i$. Given that the model in \eqref{equ:DGP_IF} is not readily identified from the data, this is not yet operational. After further derivations and assuming  $\mathbf{B} = \left[\begin{smallmatrix} b_{ii} & 0 \\ 0 & b_{jj} \end{smallmatrix}\right]$, \cite{Liang2014} transforms  \eqref{equ:IFequation_general} into  a workable formula made of empirical moments

\begin{align} \label{equ:IFequation_EconNotation}
T_{j \rightarrow i} = \frac{\sigma_{i,i} \, \sigma_{i,j} \, \sigma_{j, \Delta i} - \left(\sigma_{i,j}\right)^2 \, \sigma_{i, \Delta i}}{\left(\sigma_{i,i}\right)^2 \, \sigma_{j,j} - \sigma_{i,i} \, \left(\sigma_{i,j}\right)^2} 
\end{align}
where $\sigma_{i,i}$ is the variance of $i$,  $\sigma_{ij,}$ the covariance between $i$ and $j$, and  $\sigma_{i, \Delta i}$ is the covariance between $i$ and the $k^{th}$ difference of $i$ (we follow \cite{Liang2015} and set $k = 1$). However, the validity of this appealing formula rests on the seemingly technical assumption of a diagonal $\mathbf{B}$. Our point is that this assumption is far from merely technical and very frequently wrong. Motivating the diagonal $\mathbf{B}$ among other assumptions, \cite{Liang2014} states
\begin{quote}
\textit{Since the dynamics is unknown, we first need to choose
a model. As always, a linear model is the natural choice, at
least at the initial stage of development.}
\end{quote}
The problem is that  \textbf{(i)} of course, we must choose an empirical model, but we will try to avoid those that the data blatantly reject, \textbf{(ii)} $\mathbf{B}$ being diagonal has nothing to do with linearity, and \textbf{(iii)} there is no further potential "development" possible without this assumption, which, in effect, assumes the causal problem away. As a result, whenever the diagonal $\mathbf{B}$ is violated by the data, IFs -- as currently used in empirical studies -- provide spurious causality. 

We now use a very popular framework, from macroeconomics, to think more clearly about $\mathbf{B}$. \cite{Liang2008}'s flow of simplifications and assumptions makes his once sophisticated \eqref{equ:DGP_IF} collapse to that of a bivariate VAR with 1 lag
\begin{equation}\label{var_original}
\left[\begin{array}{l}
X_{1,t} \\
X_{2,t}
\end{array}\right]=\underbrace{\left[\begin{array}{l}
c_{1} \\
c_{2}
\end{array}\right]}_{\boldsymbol{c}}+\underbrace{\left[\begin{array}{ll}
a_{11}^{1} & a_{12}^{1} \\
a_{21}^{1} & a_{22}^{1}
\end{array}\right]}_{\boldsymbol{A}}
\left[\begin{array}{l}
X_{1,t-1} \\
X_{2,t-1}
\end{array}\right]+\underbrace{\left[\begin{array}{ll}
b_{11} & b_{12} \\
b_{21} & b_{22}
\end{array}\right]}_{\mathbf{B}}
\left[\begin{array}{l}
\varepsilon_{1 t}\\
\varepsilon_{2 t}
\end{array}\right].
\end{equation}
The seemingly innocuous assumption is much less so within a statistical framework relating assumptions directly to observable quantities. Indeed, the uncorrelatedness of $W$'s, which here translates to that of $\epsilon$, combined with $b_{21}=b_{12}=0$ have a very stark implication. Let the number of endogenous variables be $M$ and the number of lags $P$. Provided imposing $M=2$ and $P=1$ is reasonable \eqref{equ:IFequation_EconNotation} is valid if and only if regression residuals $\left[\begin{smallmatrix}
u_{1 t}\\
u_{2 t}
\end{smallmatrix}\right] = \left[\begin{smallmatrix} b_{11} & b_{12} \\ b_{21} & b_{22} \end{smallmatrix}\right] \left[\begin{smallmatrix}
\varepsilon_{1 t}\\
\varepsilon_{2 t}
\end{smallmatrix}\right]$ are \textit{not} cross-correlated.  This is easily testable: one needs to estimate equations 1 and 2 separately by least squares, collect the residuals and calculate their correlation $\rho_u$. If the latter is different from 0 (and this could be formally tested with a t-test), then \cite{Liang2008}'s simple formula does not apply. While $\rho_u=0$ might be plausible in continuous time (or anything near it), this is a monumental stretch  for data sampled at the yearly frequency (like in SMCGL). Fortunately, unlike true structural causality, the data can directly inform us on whether \cite{Liang2014}'s formula is valid or not for specific pairs of time series.

\subsection{Acknowledging the Identification Problem: Vector Autoregressions}

This exposition follows closely \cite{VARCTIC}.  In time series analysis, the "identification" problem originates from simultaneity in the data. We can learn whether 
$$X_{t-1} \rightarrow Y_{t} \quad \text{or} \quad Y_{t-1} \rightarrow X_{t}$$
is more plausible. This is predictive causality in the sense of \cite{granger1969}. However, the data itself cannot discriminate between
$$X_{t} \rightarrow Y_{t} \quad \text{and} \quad X_{t} \leftarrow Y_{t}.$$
In words, a correlation between $X_t$ and $Y_t$ can be generated by two different causal structures. \cite{Liang2008}'s solution is to assume such relationships do not exist --- yet, they do. Within a VAR, the problem boils down to the need for identifying $C$ in 
\begin{align}\label{struct_VAR}
C\boldsymbol{y}_t = \Psi_0 + \sum_{p=1}^{P}{\Psi_p}\boldsymbol{y}_{t-p} + \boldsymbol{\varepsilon}_t  ,
\end{align}
where $\boldsymbol{y}_t$ is an $M$ by 1 vector -- meaning the dynamic system incorporates $M$ variables. $\Psi_p$'s parameterizes how each of these variables is predicted by its own lags and lags of the $M-1$ remaining variables. $P$ is the number of lags being included. The matrix $C$ characterizes how the $M$ different variables interact contemporaneously --- e.g., how total forcing affects GMTA within the \textit{same} year (a time unit $t$ in SMCGL). Finally, the \textit{structural} disturbances are mutually \textit{uncorrelated} with mean zero:	
\[
\boldsymbol{\varepsilon}_t = \left[ \varepsilon_{1,t},\enskip ... \enskip , \varepsilon_{M,t} \right]  
~ \sim ~ \, N \left (0 , ~ I_M \right ).
\]
Equation (\ref{struct_VAR}) is the so-called structural form of the VAR, which \textit{cannot} be estimated because $C$ is not identified by the data. SMCGL uses formula \eqref{equ:IFequation_EconNotation} which implicitly assumes a constrained version of \eqref{struct_VAR} with $M=2$, $P=1$, and, most importantly, $C$ being a diagonal matrix. The validity of their analysis hinges upon those constraints not being rejected by the data. In section \ref{sec:empirics}, we find that the data disagrees with at least two of them.

Equation \eqref{struct_VAR} is a structural model which can be used to answer causal questions directly. However, the elements of $C$ are not plain regression coefficients, and cannot be estimated as such --- they would be biased. It does not mean that they do not exist. The implications of their existence can best be understood by looking at an estimable "reduced-form" VAR
\begin{align}\label{rf_VAR}
\boldsymbol{y}_t = \boldsymbol{c} + \sum_{p=1}^{P}{\Phi_p}\boldsymbol{y}_{t-p} + \boldsymbol{u}_t  ,
\end{align}
where $\boldsymbol{c} =  C^{-1} \Psi_0$ and $\Phi_p = C^{-1} \Psi_p$ are both regression coefficients obtained by running least squares separately on each equation. $\boldsymbol{u}_t$ are now regression residuals
\[
\boldsymbol{u}_t = \left[ u_{1,t},\enskip ... \enskip , u_{M,t} \right]  
~ \sim ~ \, N \left (0 , ~ \Sigma_u \right )
\]
\textit{which will be cross-correlated} if the true $C$ is \textit{not} diagonal. As mentioned earlier, \cite{Liang2014}'s simplifying assumptions translate into $\Sigma_u=C^{-1'}C^{-1}$ being diagonal which is often at odds with the data. All the parameters of \eqref{rf_VAR} can be estimated with traditional methods, but the model is not "structural" and cannot be used for causal inference. Structural VARs, which aim at uncovering "structural" causality (instead of predictive causality à la \citealt{granger1969}) acknowledge $\Sigma_u$ being non-diagonal and provide ways to obtain $C$. As a byproduct, they can procure valid measures of information flow. 

The raw material of causal measures are $\boldsymbol{\varepsilon}_t$, the structural disturbances entering the systems. However, those, like structural causality, are not directly extractable from the data: we only have $\boldsymbol{u}_t$ and translating those back to $\boldsymbol{\varepsilon}_t$ necessitates $C$. The latter is not directly attainable, but can be retrieved using the mapping $\Sigma_u=C^{-1'}C^{-1}$. In words, this means the covariance matrix of regression residuals from \eqref{rf_VAR} can be used as raw material to retrieve the "structural" $C$. Mechanically, the identification problem emerges because there are many $C$'s satisfying $\hat{\Sigma}_u=C^{-1'}C^{-1}$ --- numerous causal structures deliver the same residuals' cross-correlations.

The strategy we opt for is the traditional Choleski decomposition of $\hat{\Sigma}_u$. This is one of many identification strategies for the VAR \citep{kilian2017svar}. But among the catalog of methodologies, the Choleski decomposition is certainly popular (if not the most popular) in applied work and is simple to implement.  Mechanically, it provides a lower-triangular matrix $C$, satisfying $\hat{\Sigma}_u=\mathbf{B}'\mathbf{B}$ where $\mathbf{B}$ is the same as from equation \eqref{var_original}, but with dimensions $M \times M$. Its purpose is to transform  cross-correlated regressions residuals $u_t$ (equation \eqref{rf_VAR}) into uncorrelated structural shocks $\boldsymbol{\varepsilon}_t$ (equation \eqref{struct_VAR}). This is done by reversing the relationship $\boldsymbol{u}_t=C\boldsymbol{\varepsilon}_t$.

Uncorrelatedness is essential to study how GMTA responds to a given forcing, \textit{keeping everything else constant}. Such a causal claim would be impossible when considering an impulse from correlated residuals $u_t$ \textit{as those always co-move.} A Choleski decompostion of $\Sigma_u$ is \textit{one way} to transform the observed  $u_t$ into the unobserved fundamental shocks $\boldsymbol{\varepsilon}_t$. The assumption underlying such an approach to orthogonalization is a causal ordering of shocks. The ordering restricts how variables interact with each other \textit{within the same year}, conditional on the previous state of the system. In our bivariate setup, ordering forcing $j$ after GMTA implies that forcing cannot impact GMTA within the same year. Ordering GMTA after a given forcing implies that GMTA cannot impact forcing $j$ within the same year. SMCGL implicitly assumes both restrictions at the same time, which results in a rejected overidentified model. In contrast, when the model is just identified (when only one restriction is imposed), this choice cannot be validated by the data itself as it does not alter the likelihood. 

When revisiting SMCGL's empirical work, we consider both orderings and document how sensitive conclusions are to that necessary choice.\footnote{Of course, there are identification schemes outside of the family of "orderings" obtained by Choleski decomposition, but those are beyond the scope of this paper -- and unnecessary to make our main point.}  There are cases where the sign of net IF (a qualitative notion) between $j$ and GMTA depends on the ordering choice, and cases where it does not. For instance, we will find that whatever is assumed about $C$ in a proper VAR system, total forcing appears to be causing GMTA much more than the reverse. In other cases, like $CO_2$-induced radiative forcing, this cannot be simply ruled out by the data.



\subsection{An Adequate Measure of IF Based on the VAR}

In a complete multivariate system like a VAR, the errors of the $h$-steps ahead forecast $y_{t+h,m}$ can be related back to structural shocks -- that is, the anomalies driving the dynamics of the system. For instance, we can compute the share of the forecast error variance of GMTA 10 years from now that is attributable to $CO_2$ anomalies. Intuitively, if $CO_2$ is causing GMTA to increase, its exogenous impulses should be an important driver of GMTA's variance rather than GMTA anomalies themselves -- a high information flow/transfer between the two variables. Accordingly, VAR forecast errors are 
$$
\boldsymbol{u}_{t+h} = \boldsymbol{y}_{t+h}-\hat{\boldsymbol{y}}_{t+h}=\sum_{h'=0}^{h-1} \Theta_{h'} \boldsymbol{\varepsilon}_{t+h-h'}
$$
where $\Theta_{h}$ is a function of $\Phi_p$'s and $C$. See \cite{kilian2017svar} for further details. The forecast error variance decomposition (FEVD) of the whole system at horizon $h$ can be analytically calculated using the entries of matrix $\Theta_{h'}$. Precisely, 
$$
\begin{aligned}
\operatorname{MSPE}_{h} = E(\boldsymbol{u}_{t+h} \boldsymbol{u}{t+_h}') = \sum_{h'=0}^{h-1} \Theta_{h'} \Theta_{h'}^{\prime} \; .
\end{aligned}
$$
An information flow "share" of $i$ for $j$ can be characterized by the share of forecast error variance of variable $j$ attributable to structural shocks of $i$.\footnote{For a discussion on how to think about "shocks" in a physical system, see \cite{VARCTIC}.} While those measures can be assessed for any horizon $h$ (which can contain useful information), we focus on the cumulative sum since it is a measure of total flow. For non-stationary VARs (like those of the empirical section) we use a horizon of $h=15$ years. 
In the case of stationary VAR process, FEVD measures quickly converge to their long-run value as $h$ increases. Moreover, in that case, our FEVD-based IF measures can be more directly motivated from the Wold representation of a VAR process \citep{kilian2017svar}. Finally, note that whenever the $b_{12}=b_{21}=0$ assumption is approximately true, the FEVD approach and IFs give qualitatively similar answers.

\section{Simulations}\label{sec:simuls}

This section showcases how IF can lead to a pretense of causal knowledge, with conclusions that are sometimes the exact opposite of the truth. We consider four data-generating processes (DGPs) where the true $P$ is one. The first two correspond to what is proposed in \cite{Liang2014}. The last two are generic VAR positively autocorrelated processes with differing degrees of persistence. Following the notation above (equation \eqref{var_original}):

\begin{subequations}
	\begin{empheq}[left={DGP(1) \; := \; \empheqlbrace}]{align}
			\boldsymbol{A} &= \left[\begin{array}{ll}
0.5 & 0.5 \\
0 & 0.6
\end{array}\right] , \quad  \boldsymbol{c} = \left[\begin{array}{ll}
0.1 \\
0.7
\end{array}\right] 
	\end{empheq}

	\begin{empheq}[left={DGP(2) \; := \; \empheqlbrace}]{align}
			\boldsymbol{A} &= \left[\begin{array}{ll}
-0.5 & 0.9 \\
-0.2 & 0.5
\end{array}\right] , \quad  \boldsymbol{c} = \left[\begin{array}{ll}
0 \\
0
\end{array}\right] 
	\end{empheq}

	\begin{empheq}[left={DGP(3) \; := \; \empheqlbrace}]{align}
			\boldsymbol{A} &= \left[\begin{array}{ll}
0.5 & -0.2 \\
-0.5 & 0.25
\end{array}\right] , \quad  \boldsymbol{c} = \left[\begin{array}{ll}
0 \\
0
\end{array}\right] 
	\end{empheq}

	\begin{empheq}[left={DGP(4) \; := \; \empheqlbrace}]{align}
			\boldsymbol{A} &= \left[\begin{array}{ll}
0.25 & -0.1 \\
-0.2 & 0.1
\end{array}\right] , \quad  \boldsymbol{c} = \left[\begin{array}{ll}
0 \\
0
\end{array}\right] 
	\end{empheq}
\end{subequations}

As highlighted in the previous section, IFs are calculated assuming $b_{ij}=0$. However, as will be reported and formally tested in section \ref{sec:empirics}, this is frequently not the case for most time series, especially those sampled at low frequencies. Using a controlled simulation environment, we study how IFs behave for values of $\rho_u := corr(u_t^{X_1},u_t^{X_2}) \in [-1,1]$.\footnote{Variances of $u_t^{X_1}$ and $u_t^{X_2}$ are one.} Note that IFs are invariant to $\rho_u \neq 0$ emerging from $b_{12}=0$ or $b_{21}=0$, which is obviously problematic given the causal content of $b_{ij}$.\footnote{It is important to note that while we consider cases where either $b_{12}=0$ or $b_{21}=0$, there is a continuum of possibilities between those. We do so for simplicity of exposition (it makes the problem dichotomous). Moreover, setting either $b_{12}=0$ or $b_{21}=0$ to 0 corresponds to a causal ordering which is by far the most common identification scheme used in practice -- and what we will be using in section \ref{sec:empirics}.} 

As our indicator of true underlying information flow, taking into account $\rho_u \neq 0$, we use the FEVD-based measure described earlier. Note that here, unlike the application to real data in section \ref{sec:empirics}, \textit{we know} what are the true $b_{ji}$'s. When $\rho_u \neq 0$, the correlation must be attributed to either $b_{12}$ or $b_{21}$, or a combination of both. In a bivariate setup, this amounts to setting $b_{ij}=0$, and attributing $\rho \neq 0$ to $b_{ji}$. Hence, it is possible to tell when standard IFs conclude falsehoods, because in simulations $b_{12}$ and $b_{21}$ are known. 

In terms of notation, $\Upsilon^{i,j}_{i \rightarrow j}$ means the FEVD share at horizon $h = 10$ with $i = 1, 2$ and $i \neq j$. The superscripts in $\Upsilon^{i,j}_{i \rightarrow j}$ determine the \textit{true} ordering, hence $i$ ordered before $j$. The subscripts indicate that we plot the contribution of variable $i$ to the forecast error variance of variable $j$ at horizon $h = 10$. The simulations have shown that $h = 10$ is sufficient for convergence.

\begin{figure}[t!]
	\begin{center}
	\begin{subfigure}[t]{0.5\textwidth}
		\includegraphics[trim={0mm 0mm 0mm 10mm},clip,width=\textwidth]{{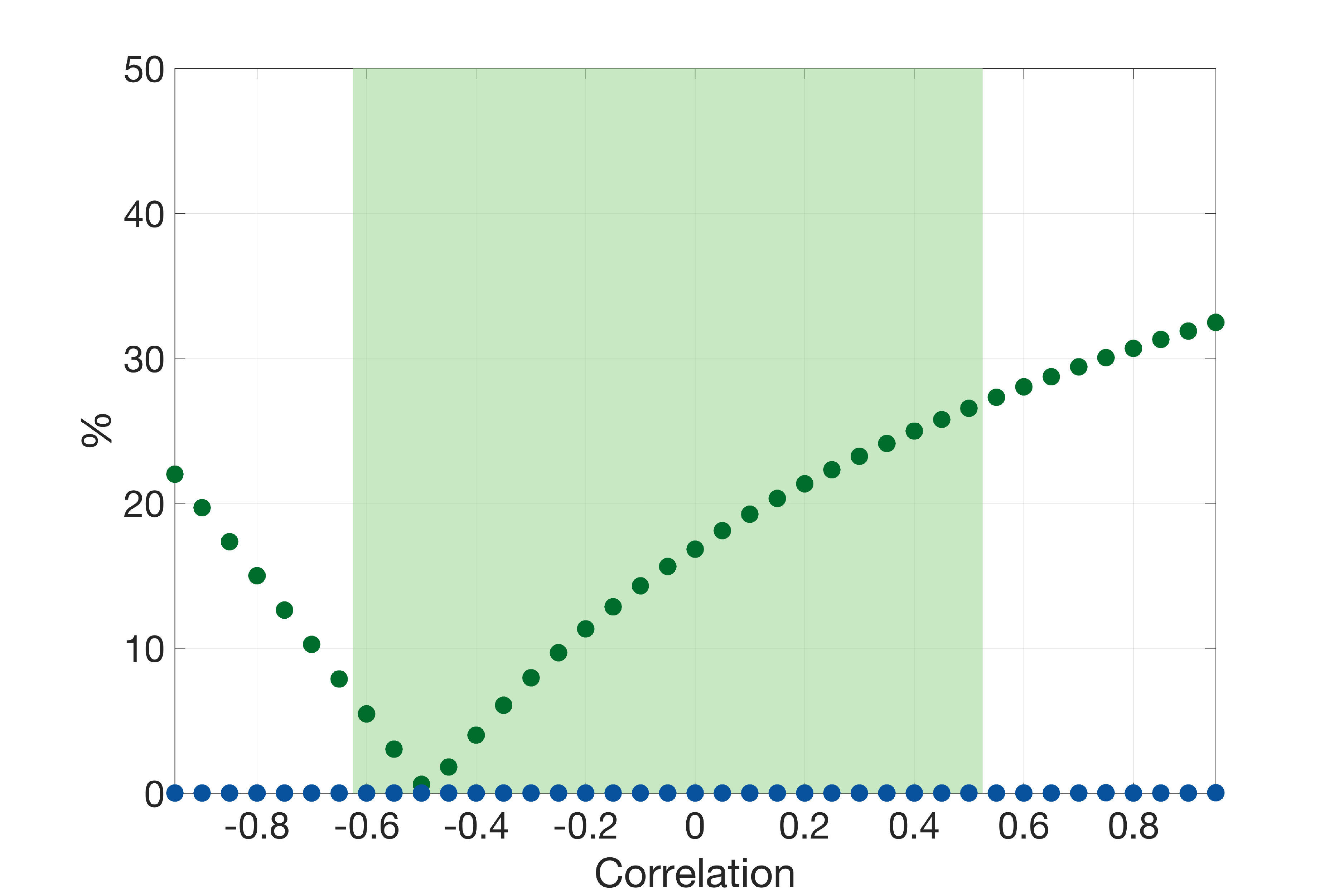}} 
		\caption{DGP(1)} 
	\end{subfigure}%
	\begin{subfigure}[t]{0.5\textwidth}
		\includegraphics[trim={0mm 0mm 0mm 10mm},clip,width=\textwidth]{{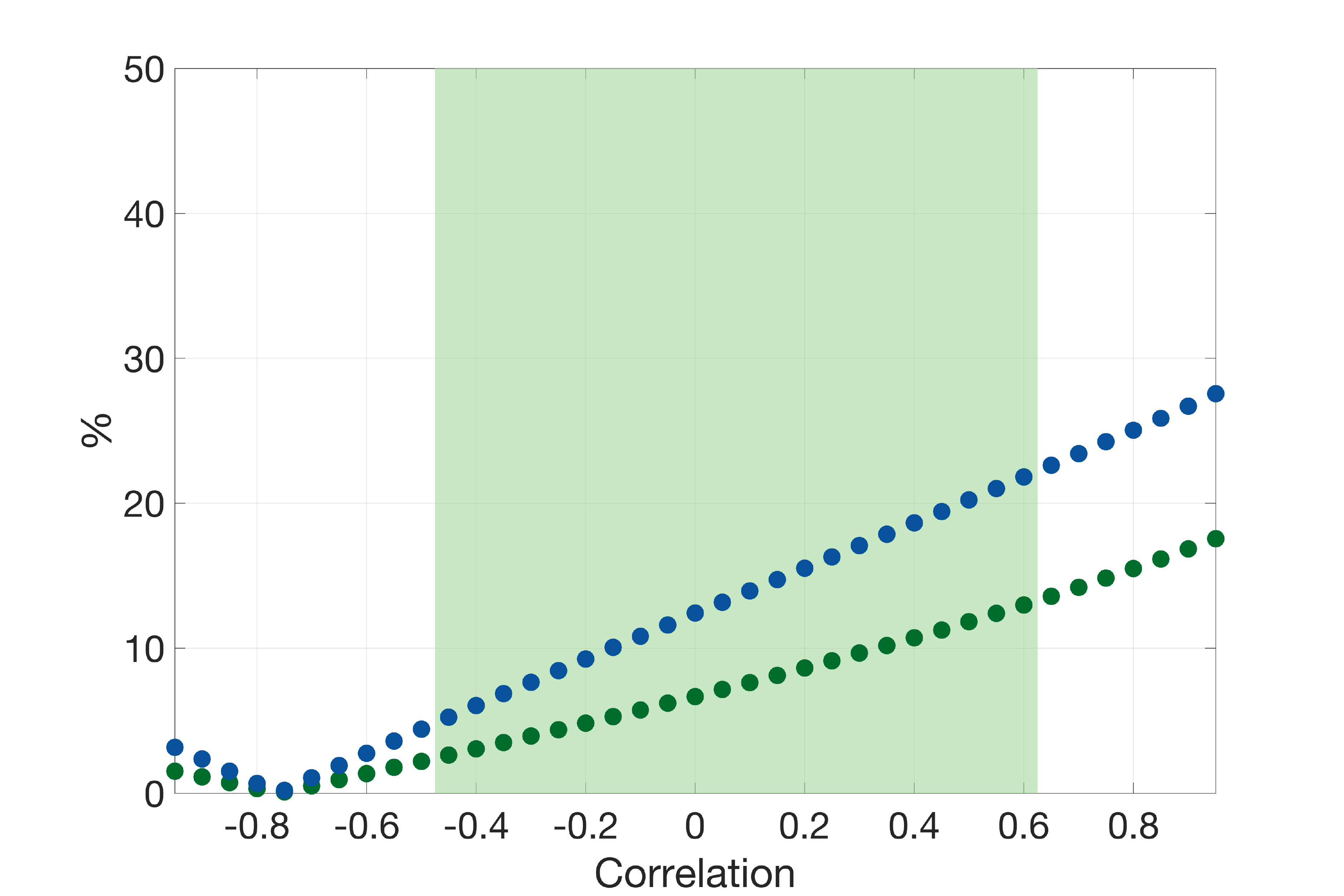}} 
		\caption{DGP(2)} 
	\end{subfigure}
\vskip 10pt
	\begin{subfigure}[t]{0.5\textwidth}
		\includegraphics[trim={0mm 0mm 0mm 10mm},clip,width=\textwidth]{{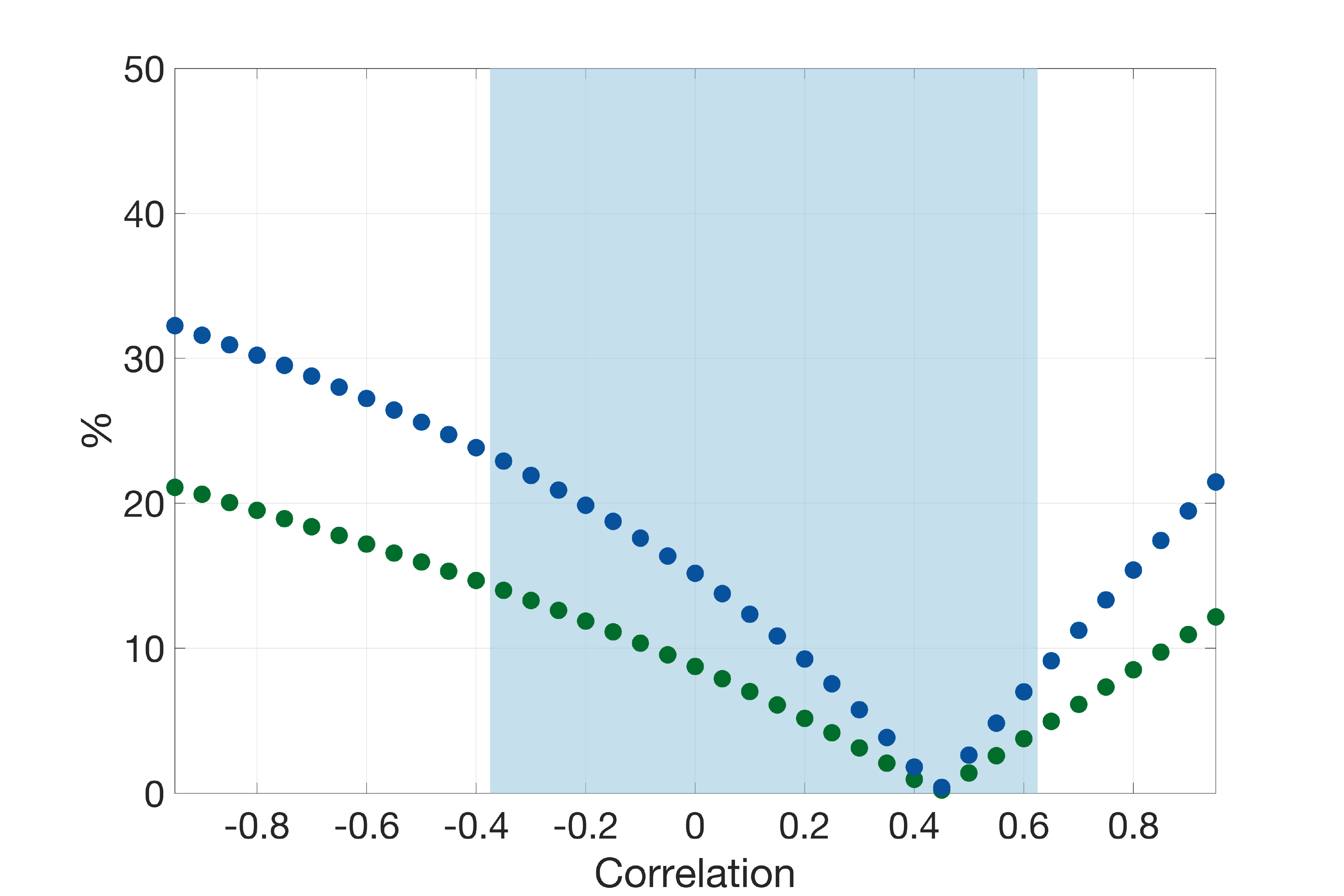}} 
		\caption{DGP(3)} 
	\end{subfigure}%
	\begin{subfigure}[t]{0.5\textwidth}
		\includegraphics[trim={0mm 0mm 0mm 10mm},clip,width=\textwidth]{{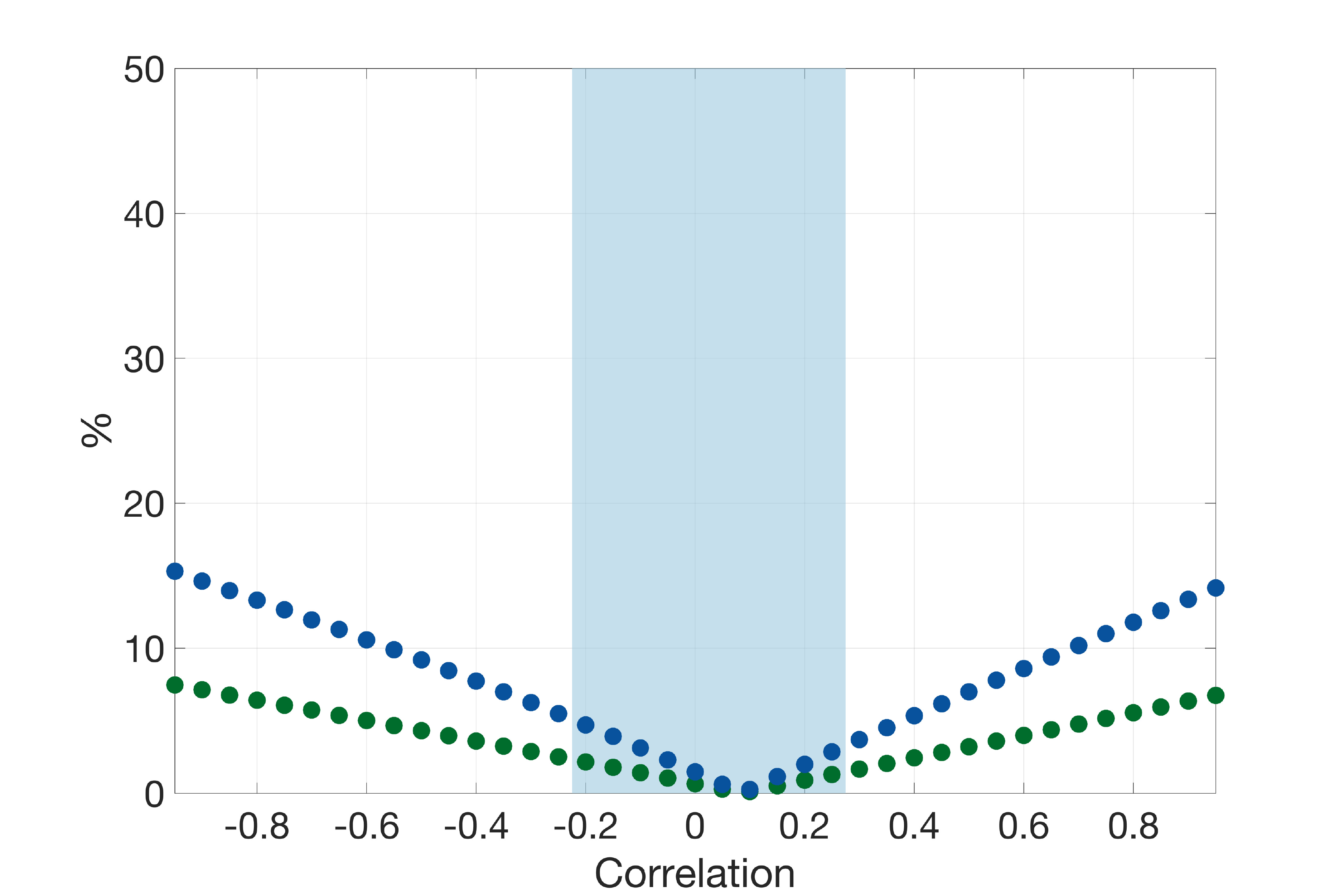}} 
		\caption{DGP(4)}
	\end{subfigure}

	\begin{subfigure}[t]{\textwidth}
	\centering
	\vspace*{0.5cm}	
		\includegraphics[trim={0mm 0mm 0mm 0mm},clip,scale=0.4]{{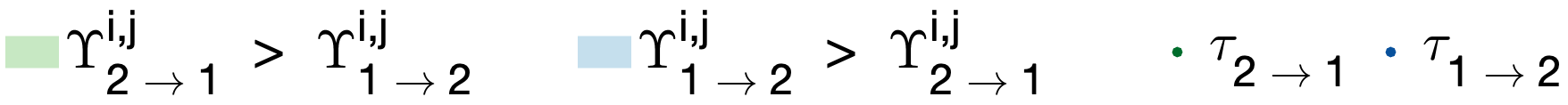}} \vskip 5pt
	\end{subfigure}
	\vskip 5pt
	\end{center}
	\noindent \scriptsize \textbf{Notes}: Color specification: the \emph{light blue} area shows the region, in which FEVDs suggest $\Upsilon^{i,j}_{1 \rightarrow 2}$ $>$ $\Upsilon^{i,j}_{2 \rightarrow 1}$, regardless of the ordering. Similarly, the \emph{light green} area shows the region, in which FEVDs suggest $\Upsilon^{i,j}_{2 \rightarrow 1}$ $>$ $\Upsilon^{i,j}_{1 \rightarrow 2}$ -- independent of the ordering. The \emph{white/non-colored} areas shows those regions, for which the ordering of $X1$ and $X2$ does matter. An unambiguous determination of $\Upsilon^{i,j}_{1 \rightarrow 2}$ $>$ $\Upsilon^{i,j}_{2 \rightarrow 1}$, respectively $\Upsilon^{i,j}_{2 \rightarrow 1}$ $>$ $\Upsilon^{i,j}_{1 \rightarrow 2}$, is not possible.
	\vskip 5pt
		\caption{Ranking of NIFs ($\tau_{i \rightarrow j}$) vs Ranking of FEVDs ($\Upsilon^{i,j}_{i \rightarrow j}$) \\ Different Levels of Correlation; Horizon $h = 10$}
	\label{fig:DIFFs_TauPlot_matching}
\end{figure}

Figure \ref{fig:DIFFs_TauPlot_matching} plots the absolute normalized information flows (NIF), $\tau_{i \rightarrow j}$  for DGP(1) through DGP(4), with varying $\rho_u$. A first observation is that $\tau_{i \rightarrow j}$ often varies significantly with $\rho_u$, even when the very formula underlying it assumes $\rho_u=0$. The relative strength of $\tau_{1 \rightarrow 2}$ and $\tau_{2 \rightarrow 1}$ can easily collapse to 0 or be much higher than what IFs report for $\rho_u=0$.

The fact that IFs vary with $\rho_u$ could give false hopes that they account for simultaneous relationships. We conduct a simple exercise to show it does not. An interesting empirical question is whether $i$ is causing $j$ more than $j$ is causing $i$, which is the appealing promise of IFs. This moves beyond testing for Granger Causality and aims at quantifying the flow of information.  The shaded green region corresponds to the values of $\rho_u$ for which either the true underlying causality is that $X_2$ causes  $X_1$ more than the reverse, \textit{irrespective} of whether 
it emerges from either $b_{12}=0$ or $b_{12}=0$. For those values of $\rho_u$, when the true ordering is unknown, the \textit{qualitative} conclusion about the sign of the net causality flow does not hinge on knowledge of $b_{ij}$. Analogously, the blue shade represents values of $\rho_u$ for which $X_1$ causing  more $X_2$ is unanimous among $b_{ij}$ configurations. White regions are values of $\rho_u$ where conclusions about the sign of net causality flow cannot be determined from the data, i.e., the unknown $b_{ij}$ is necessary to settle. In other words, given the data available to the modeler, $X_2$ causing more $X_1$ and vice versa are both equally likely, and sorting it out decisively implies making a successful guess on the true value of $b_{ij}$.

If IFs were correctly calibrated, the green dotted line should only be above the blue in the shaded green regions, and the dotted blue line above the green one in blue regions. Figure \ref{fig:DIFFs_TauPlot_matching} clearly demonstrates this not to be the case. For every DGP, there is a substantial range of values (white spaces) for which IFs clearly conclude $\tau_{i \rightarrow j} > \tau_{j \rightarrow i}$ when there is one pair of $(b_{ij},b_{ji})$ out of two \textit{for which the reverse is true}. To address concerns about an horizon mismatch $h$ between IFs and FEVD, Figure \ref{fig:DIFFs_TauPlot_matching_h2} in Appendix \ref{sec:add_simulations} shows results for $h = 2$. As it turns out, things worsen for IFs (the blue and green regions shrink) since the simultaneity problem is less diluted in dynamics at short horizons.



In sum, those simulations (largely inspired by \cite{Liang2015} DGPs) show two things. First, the tractable IFs (from formula \eqref{equ:IFequation_EconNotation}) are functions of $\rho_u$ even though they assume it to be zero. This compromises any statement on the strength of causal links. Second, for any DGP, there is an underlying pair $\left(b_{ij},b_{ji}\right)$ for which IFs' conclusion about the net causal flow is the opposite of reality. This is not only an artifact of large $|\rho_u|$, as exemplified by DGP(4).

\section{GMTA and Radiative Forcing Revisited}\label{sec:empirics}

In this section, we first revisit SMCGL's application of IFs to GMTA and forcings. Second, we look at what lies behind opaque FEVD measurements by reporting impulse response functions, and computing the transient climate response implied by simplistic bivariate VARs. 

Global warming generated by man-made forcing is the prevalent generalization of the notion \emph{climate change}. Numerous researchers have dedicated their works to this very relationship between anthropogenic forcing and constituents of global climate \citep{Hansen2006,AndrewsEtAl2010,LiEtAl2013,notz2016observed} and the socio-economic consequences thereof \citep{Nord14,ipcc2019}. Despite overwhelming evidence for anthropogenic forcing being the main driver of global climate change \citep{ipccGlobWarm2018}, scientists have also observed that especially since the turn of the millennium, global temperature has plateaued despite ever-rising greenhouse gases and contrary to projections from key climate models.\footnote{Resolving this puzzle has led to reevaluate the role of oceans in the interplay of radiative forcing and the climatic response \citep{Tollefson14,Marotzke2015}.} 


IFs hinge exactly on these bivariate relationships, which are attractive in their clarity, but are certainly a stark oversimplification of a complex system. Nevertheless, to give empirical content to our critique of the methodology, we study the relationship between GMTA and 7 forcings from SMCGL using both IFs and our FEVD-based remedy.  The sample of annual means ranges from 1850 - 2005.\footnote{Data on the Pacific Decadal Oscillation (PDO) ranges from 1900 - 2005.}  We follow SMCGL and take our data from therein referenced data providers. Table \ref{tab:NIF_FEVD} summarizes the results.\footnote{The sample was restricted to 1850-2005 to match that of SMCGL. Table \ref{tab:TCR_18502017} reports results extending the sample to 2017.}

In Table \ref{tab:NIF_FEVD} we report estimated correlation between residuals of the bivariate VAR(1) implied by IFs. When forcing $P=1$, in 5 cases out of 7, the null that $\hat{\rho}_u^{i,GMTA}=0$ is rejected at least at the 10\% level. When choosing $P$ with Bayesian Information Criterion (BIC), only $\hat{\rho}_u^{\text{Solar},GMTA}=0$ cannot be rejected. Hence, as repeatedly mentioned in the text, IFs assume something that can be, and is rejected by the data. Moreover, in the light of simulations carried earlier, the qualitative and quantitative insights from IFs are often spurious under such conditions.

\begin{table} [h!]
\renewcommand{\arraystretch}{1.2}
	\caption{Empirical Results for the Bivariate Relationship \\ Between Various Forcings and GMTA}
	\vspace*{-0.5cm}
	\label{tab:NIF_FEVD}
	\begin{center}
		{ \scriptsize
		\begin{adjustbox}{max width=\textwidth}
			\begin{tabular}{l c cc  c ccccc}
				\toprule \toprule \addlinespace[5pt]
				
				&   \multicolumn{1}{c|}{\multirow{3}{*}{\textbf{Correlation}}} & \multicolumn{2}{c}{\multirow{2}{*}{\textbf{\makecell{ Normalized IF \\ [3pt] (IF $\times$ 100) }}}}& \multicolumn{6}{|c}{\textbf{FEVD}} \\	
				\cmidrule(lr){5-10}
				
				&   \multicolumn{1}{c|}{}                &  & & \multicolumn{1}{|c}{\multirow{2}{*}{\textbf{\makecell{Lags \\ [3pt] ($P$)} }}} & \multirow{2}{*}{\textbf{\makecell{Correlation \\ of \\ Residuals ($\rho_u$)}}} & \multicolumn{2}{c}{\textbf{Ordering:} $i$, GMTA} & \multicolumn{2}{c}{\textbf{Ordering:} GMTA, $i$} \\
				\cmidrule(lr){3-4} \cmidrule(lr){7-8} \cmidrule(lr){9-10}

				&	 &   \multicolumn{1}{|c}{$i$ $\rightarrow$ GMTA} & \multicolumn{1}{c|}{GMTA $\rightarrow$ $i$} & & &
				${i \rightarrow \text{GMTA}}$ & ${\text{GMTA} \rightarrow i}$ &
				${i \rightarrow \text{GMTA}}$ & ${\text{GMTA} \rightarrow i}$ \\  [3pt]
				
\cmidrule(lr){1-1} 	\cmidrule(lr){2-2} \cmidrule(lr){3-3} \cmidrule(lr){4-4}  \cmidrule(lr){5-5} \cmidrule(lr){6-6}  \cmidrule(lr){7-7} \cmidrule(lr){8-8}  \cmidrule(lr){9-9} \cmidrule(lr){10-10}  \addlinespace [5pt]

\multicolumn{1}{l|}{\multirow{2}{*}{Total Forcing}} 
& \multirow{2}{*}{0.73}  & \textbf{30.6} & \multicolumn{1}{c|}{20.8} & 

  \cellcolor{gray!15} 4 &  \cellcolor{gray!15} 0.23*** &
\cellcolor{gray!15} \multirow{1}{*}{\textbf{47.4}} & \cellcolor{gray!15} \multirow{1}{*}{13.0}  & \cellcolor{gray!15} \multirow{1}{*}{\textbf{28.0}}  & \cellcolor{gray!15} \multirow{1}{*}{25.4} \\ 

\multicolumn{1}{l|}{}&   & (15.3) & \multicolumn{1}{c|}{(11.1)} & 
  1 & 0.29*** &
\multirow{1}{*}{\textbf{51.4}} & \multirow{1}{*}{9.6} & \multirow{1}{*}{27.6}  & \multirow{1}{*}{\textbf{28.7}} 
 \\ \addlinespace[5pt]

\midrule \addlinespace [10pt]

\multicolumn{1}{l|}{\multirow{2}{*}{Anthropogenic}} 
& \multirow{2}{*}{0.86}  & \textbf{39.8} & \multicolumn{1}{c|}{-20.0} & 

  \cellcolor{gray!15} 4 & \cellcolor{gray!15} -0.19** &
\cellcolor{gray!15} \multirow{1}{*}{\textbf{6.5}} & \cellcolor{gray!15} \multirow{1}{*}{3.7}  & \cellcolor{gray!15} \multirow{1}{*}{3.8}  & \cellcolor{gray!15} \multirow{1}{*}{\textbf{13.4}}  \\ 

\multicolumn{1}{l|}{}&  & (35.7) & \multicolumn{1}{c|}{(-0.6)} & 
 1&  -0.19** &
\multirow{1}{*}{5.0} & \multirow{1}{*}{\textbf{5.8}} & \multirow{1}{*}{2.2}  & \multirow{1}{*}{\textbf{17.1}}  \\ \addlinespace[5pt]

\midrule \addlinespace [10pt]

\multicolumn{1}{l|}{\multirow{2}{*}{\makecell{CO$_2$ - ERF (W/m$^\text{2}$) \\ SMCGL}}}
& \multirow{2}{*}{0.86} &  \textbf{39.6} & \multicolumn{1}{c|}{-15.2} & 

\cellcolor{gray!15}  4 & -\cellcolor{gray!15} 0.14*  & 
\cellcolor{gray!15} \multirow{1}{*}{6.5} & \cellcolor{gray!15} \multirow{1}{*}{\textbf{8.4}}  & \cellcolor{gray!15} \multirow{1}{*}{5.6}  & \cellcolor{gray!15} \multirow{1}{*}{\textbf{17.4}}  \\ 

\multicolumn{1}{l|}{}& &  (35.1) & \multicolumn{1}{c|}{(-0.4)} & 
 1 & -0.15* &
\multirow{1}{*}{2.8} & \multirow{1}{*}{\textbf{4.7}} & \multirow{1}{*}{1.1}  & \multirow{1}{*}{\textbf{12.8}} \\ \addlinespace[5pt]

\midrule \addlinespace [10pt]

\multicolumn{1}{l|}{\multirow{2}{*}{Aerosol}} 
& \multirow{2}{*}{-0.82} &  \textbf{35.9} & \multicolumn{1}{c|}{-24.5} &
 
\cellcolor{gray!15} 4 &\cellcolor{gray!15}  -0.19** & 
\cellcolor{gray!15} \multirow{1}{*}{\textbf{2.9}} & \cellcolor{gray!15} \multirow{1}{*}{0.6}  & \cellcolor{gray!15} \multirow{1}{*}{\textbf{2.1}}  & \cellcolor{gray!15} \multirow{1}{*}{1.6} \\ 

\multicolumn{1}{l|}{}&  & (24.3) & \multicolumn{1}{c|}{(-0.4)} & 
  1 & -0.10 &
\multirow{1}{*}{3.5} & \multirow{1}{*}{\textbf{4.0}} & \multirow{1}{*}{\textbf{1.8}}  & \multirow{1}{*}{1.2}  \\ \addlinespace[5pt]

\midrule \addlinespace [10pt]

\multicolumn{1}{l|}{\multirow{2}{*}{Solar}} 
& \multirow{2}{*}{0.49} &  \textbf{13.5} & \multicolumn{1}{c|}{6.7} & 

\cellcolor{gray!15}  8 & \cellcolor{gray!15} 0.05 &
\cellcolor{gray!15} \multirow{1}{*}{\textbf{8.5}} & \cellcolor{gray!15} \multirow{1}{*}{1.6}  & \cellcolor{gray!15} \multirow{1}{*}{\textbf{6.6}}  & \cellcolor{gray!15} \multirow{1}{*}{2.5}  \\ 

\multicolumn{1}{l|}{}& &  (3.8) & \multicolumn{1}{c|}{(2.3)} & 
 1 & 0.08 &
\multirow{1}{*}{\textbf{16.6}} & \multirow{1}{*}{4.2} & \multirow{1}{*}{\textbf{12.3}}  & \multirow{1}{*}{6.8}  \\ \addlinespace[5pt]

\midrule \addlinespace [10pt]

\multicolumn{1}{l|}{\multirow{2}{*}{Volcanic}} 
& \multirow{2}{*}{0.09} &  \textbf{0.9} & \multicolumn{1}{c|}{-0.5} & 

\cellcolor{gray!15} 4 & \cellcolor{gray!15} 0.18** & 
\cellcolor{gray!15} \multirow{1}{*}{\textbf{10.9}} &\cellcolor{gray!15}  \multirow{1}{*}{0.8}  & \cellcolor{gray!15} \multirow{1}{*}{3.1}  & \cellcolor{gray!15} \multirow{1}{*}{\textbf{3.6}} \\ 

\multicolumn{1}{l|}{}& &  (0.2) & \multicolumn{1}{c|}{(-0.4)} & 
 1 & 0.20**  &
\multirow{1}{*}{\textbf{7.1}} & \multirow{1}{*}{1.4} & \multirow{1}{*}{0.6}  & \multirow{1}{*}{\textbf{3.7}}  \\ \addlinespace[5pt]

\midrule \addlinespace [10pt]

\multicolumn{1}{l|}{\multirow{2}{*}{PDO}} 
& \multirow{2}{*}{0.17} &  \textbf{-1.2} & \multicolumn{1}{c|}{-0.6} & 

\cellcolor{gray!15} 4 & \cellcolor{gray!15} 0.35*** & 
\cellcolor{gray!15} \multirow{1}{*}{\textbf{31.1}} & \cellcolor{gray!15} \multirow{1}{*}{0.9}  & \cellcolor{gray!15} \cellcolor{gray!15} \multirow{1}{*}{6.3}  & \cellcolor{gray!15} \multirow{1}{*}{\textbf{10.3}} \\ 

\multicolumn{1}{l|}{}& &  (-0.2) & \multicolumn{1}{c|}{(-0.5)} & 
1 & 0.34*** & 
\multirow{1}{*}{\textbf{9.1}} & \multirow{1}{*}{0.5} & \multirow{1}{*}{0.2}  & \multirow{1}{*}{\textbf{10.7}} \\ \addlinespace[5pt]

\midrule \addlinespace [10pt]

\multicolumn{1}{l|}{\multirow{2}{*}{\makecell{CO$_2$ (Mt/yr)}}} 
& \multirow{2}{*}{0.82} &  \textbf{37.1} & \multicolumn{1}{c|}{-4.3} & 

\cellcolor{gray!15} 2 & \cellcolor{gray!15}  -0.10 & 
\cellcolor{gray!15} \multirow{1}{*}{\textbf{8.9}} & \cellcolor{gray!15} \multirow{1}{*}{2.1}  & \cellcolor{gray!15} \cellcolor{gray!15} \multirow{1}{*}{\textbf{10.7}}  & \cellcolor{gray!15} \multirow{1}{*}{0.6} \\ 

\multicolumn{1}{l|}{}& &  (27.0) & \multicolumn{1}{c|}{(-0.0)} & 
 1  &  -0.05  & 
\multirow{1}{*}{\textbf{4.2}} & \multirow{1}{*}{0.0} & \multirow{1}{*}{\textbf{4.4}}  & \multirow{1}{*}{0.4} \\ \addlinespace[5pt]

\midrule \addlinespace [10pt]

\multicolumn{1}{l|}{\multirow{2}{*}{\makecell{CO$_2$ (W/m$^\text{2}$)}}} 
& \multirow{2}{*}{0.86} &  \textbf{39.5} & \multicolumn{1}{c|}{-14.0} & 

\cellcolor{gray!15} 4 & \cellcolor{gray!15} 0.23*** & 
\cellcolor{gray!15} \multirow{1}{*}{5.2} & \cellcolor{gray!15} \multirow{1}{*}{\textbf{16.8}}  & \cellcolor{gray!15} \cellcolor{gray!15} \multirow{1}{*}{2.9}  & \cellcolor{gray!15} \multirow{1}{*}{\textbf{4.7}} \\ 

\multicolumn{1}{l|}{}& &  (34.5) & \multicolumn{1}{c|}{(-0.3)} & 
1 & 0.07 & 
\multirow{1}{*}{1.6} & \multirow{1}{*}{\textbf{4.1}} & \multirow{1}{*}{0.9}  & \multirow{1}{*}{\textbf{1.8}} \\ \addlinespace[5pt]

\bottomrule	 \bottomrule
			\end{tabular}
			\end{adjustbox}
		}
	\end{center}
	
	\begin{spacing}{1.0} \footnotesize \noindent Notes: $i$ corresponds to the type of radiative forcing, listed in the left most column. The second column ("Correlation") documents the correlation between GMTA and variable $i$. FEVD values are taken at horizon $h = 15$, which translates into the contribution of variable $i$ in the variance of the forecast error of variable $j$ a decade and a half after the in-sample end date. Numbers in bold underline the highest absolute causal flow among a ($i$,GMTA) pair for a given measure. "*", "**", and "***" means that the null of the residuals cross-correlation of residuals is rejected at the 10\%,5\% and 1\% level respectively. 
	\end{spacing}
\end{table}

Naturally, we concentrate on FEVD results which correctly account for $\rho_u \neq 0$. However, setting $\rho_u=0$ is only one of the empirical shortcomings of the empirical IF formula --- it also sets $P$, the number of lags of each $X_t$, to be 1. Clearly, it is empirically plausible that $X_{i,t-2}$ or $X_{j,t-4}$ may have an impact on $X_{i,t}$ beyond what is channeled by single year lags ($X_{i,t-1}$ and $X_{j,t-1}$) . In other words, $P=1$ is extremely restrictive on climatic dynamics. When choosing $P$ with BIC, results align better with prior scientific knowledge. Nonetheless, for the sake of completeness, we report both results ($P=1$ and $P=P^*$, with $P^*$ being BIC's choice).

With $P=P^*$, the fact that total forcing causes GMTA more than the reverse is without appeal. Nevertheless, the quantitative answer is, again, highly dependent on the ordering choice. After 15 years, total forcing anomalies are responsible for explaining between 28.0\% and 47.4\% of that of GMTA -- depending on the preferred ordering. The net causal flow being higher from aerosol and solar to GMTA are also unanimous, but much smaller. Indecisive results are reported for Anthropogenic, Volcanic and PDO. Overall, Table \ref{tab:NIF_FEVD} suggests that the data itself does not support the strong qualitative conclusions of SMCGL for their CO$_2$ measure and Anthropogenic. 

When $P$ is forced to one, as in IFs, inconclusive results are reported for total forcing, aerosol, volcanic and PDO. $P=1$ specifications, irrespective of the ordering\footnote{There are other identification schemes which cannot be cast as "orderings". That is, there are rotations of $\boldsymbol{u}_t$ (even when $\rho_u=0$) giving different structural shocks. Hence, while the two orderings span a lot of possibilities (and those traditionally considered first in practice), they do represent the universe of rotations of $\boldsymbol{u}_t$ into $\boldsymbol{\epsilon}_t$.}, conclude that GMTA is causing more $CO_2$ and Anthropogenic than the reverse. Only solar forcing results are unanimous, and in line with what climatic common wisdom suggests. Note that it also agrees with results from original IFs, which is not surprising given that $\hat{\rho}_{Solar,GMTA}$ is in the close vicinity of 0. However, choosing a proper $P$ nearly doubles the share of forecast errors attributable to solar forcing. 

\subsection{What's up with CO$_2$?}

Results for CO$_2$ are rather surprising. Irrespective of the ordering, GMTA is reported to cause CO$_2$ more than the reverse, a finding contradicting SMCGL's results and common wisdom.\footnote{\cite{Hansen2006} states that global warming started to accelerate not prior to the 1970s. Only after 1975 did the global temperature increase by approximately 0.2$^\circ$C per decade.} However, SMCGL's CO$_2$ measure exhibit ill dynamic behavior that cannot possibly be that of a natural quantity. This becomes obvious when \href{https://drive.google.com/file/d/1yhaJi92dvY_Lax0H5BFIzJeNc517Gn44/view?usp=sharing}{plotting} their CO$_2$-ERF measure in first difference: it evolves according to a series of dichotomous jumps. This CO$_2$-ERF series, in which ERF stands for effective radiative forcing, is a concept presented in \cite{MyhreEtAl2013}. The series itself is based on  \cite{EtminanEtAl2016}.

Given the strange jumping behavior of the CO$_2$-ERF series, and to make our calculations comparable to other findings in the literature \citep{BrunsEtAl2020,MontamatStock2020}, we derive RF from CO$_2$ concentration as follows:
we use the well established \cite{Meinshausen2017} data set on annual global means of CO$_2$ concentration, measured in parts per million (\textit{ppm}). We follow \cite{MyhreEtAl2013} and transform the increase in CO$_2$ concentration in year $t$ measured in \textit{ppm} relative to the concentration in a given base year, CO$_{2,base}$, into radiative forcing, $RF^{\text{CO}_2}_t$ -- measured in \textit{W/m$^\text{2}$} -- as follows:
$RF^{\text{CO}_2}_t = 5.35 \times \text{ln}\left(\text{CO}_{2,t} / \text{CO}_{2,base} \right)$. 
We use 1850 as base year following \cite{BrunsEtAl2020}.

As it turns out, considering this less contentious CO$_2$ series does not resolve the apparently counterintuitive finding that GMTA explains a larger portion of the forecast error variance of CO$_2$ than vice versa. Such a finding has also been reported using a different methodology in \cite{koutsoyiannis2020atmospheric}. We explore a last avenue, that of using annual CO$_2$ emissions rather than $RF^{\text{CO}_2}_t$. This last attempt is successful in reconciling the FEVD approach with the traditional wisdom that CO$_2$ is causing GMTA "more" than the reverse. This finding is independent of the ordering choice.  



In sum, \textit{based on this particular time series evidence}, the causal link between certain forcings and GMTA remains disputable. What is less disputable are the effects of total forcing, CO$_2$ emissions, and solar forcing which all explain an important share of GMTA anomalies independently of arbitrary ordering preferences. Nonetheless, we see such analyses as rather primitive and potentially misleading. For instance, $X$ causing ''more'' $Y$ does not mean that the reverse causality is not quantitatively important, or climatologically relevant. We now turn to a more promising way to extract meaning out of selected bivariate VARs.


\subsection{Impulse Response Functions} \label{sec:empirics_IRFs}
\begin{figure}[h!]
	\caption{IRFs: Cumulative Emissions and Global Temperature}
	\begin{center}	
	\begin{subfigure}[t]{0.495\textwidth}
		\includegraphics[trim={0mm 0mm 0mm 10mm},clip,width=\textwidth]{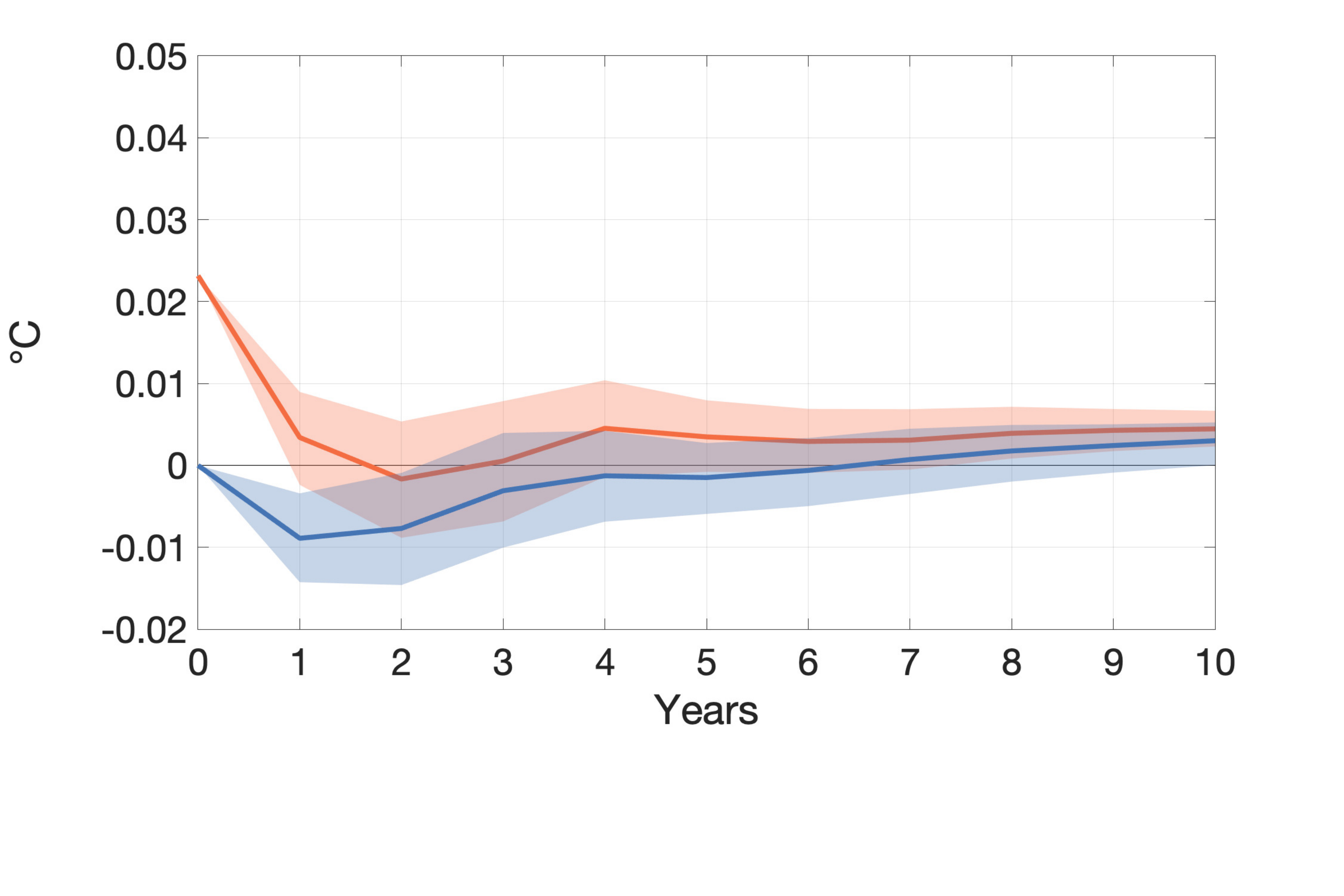} 
		\vskip -25pt
		\caption{CO$_2$ $\rightarrow$  GMTA}
	\end{subfigure}
	\begin{subfigure}[t]{0.495\textwidth}
		\includegraphics[trim={0mm 0mm 0mm 10mm},clip,width=\textwidth]{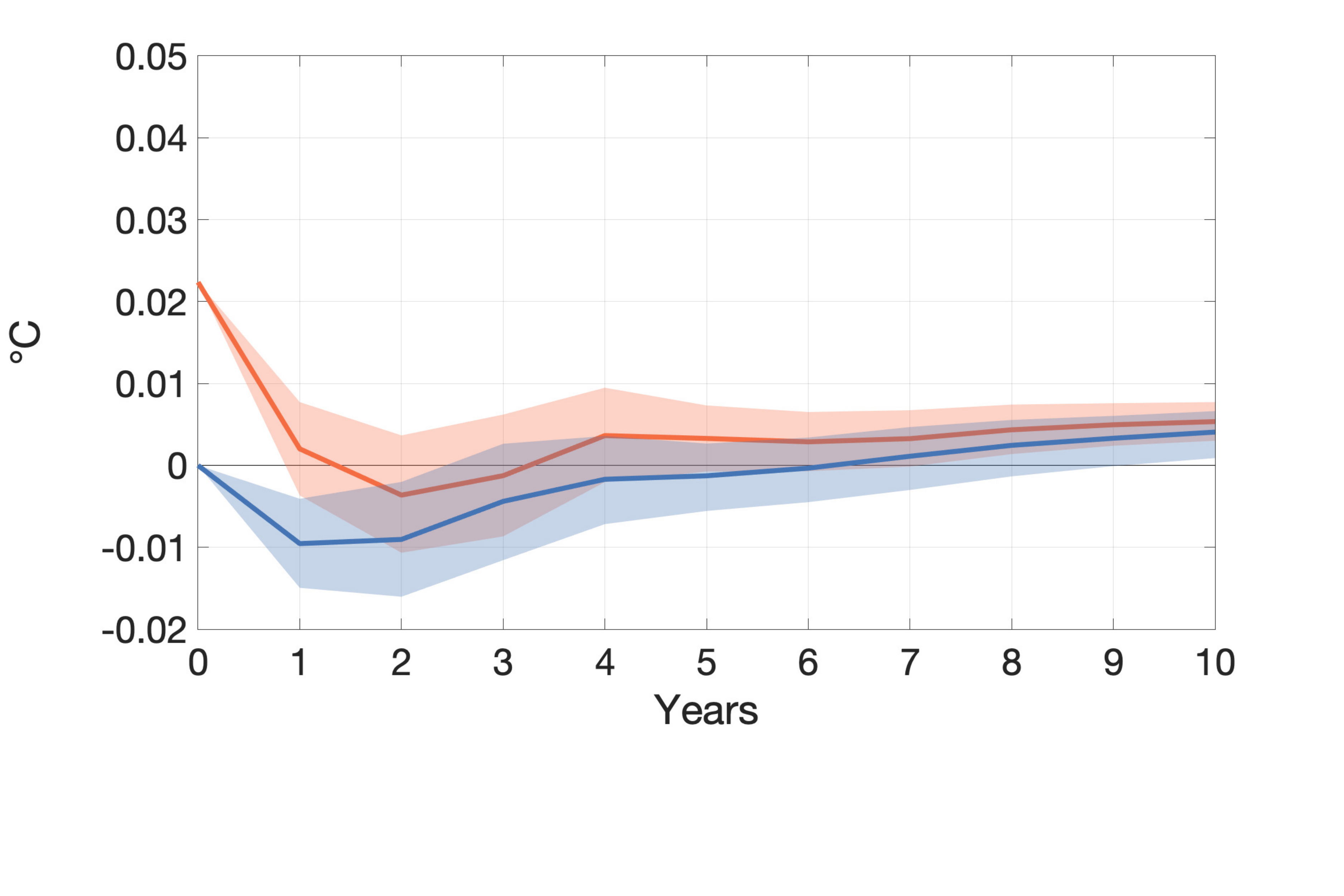} 
		\vskip -25pt
		\caption{CO$_2$ $\rightarrow$  GMTA (including trend)}
	\end{subfigure}
	\begin{subfigure}[t]{\textwidth}
	\centering
	\vspace*{0.15cm}	
		\includegraphics[trim={0mm 0mm 0mm 0mm},clip,scale=0.3]{{Figures/Replication_IRFs/Replication_IRFs_LEGEND.png}} \vskip 5pt
	\end{subfigure}
		\end{center}
	\label{fig:Replication_IRFs}
	\scriptsize Notes: We show impulse response functions from bivariate Vector Autoregressions of \emph{GMTA} and CO$_2$, and Total Forcing respectively. The right column includes a time trend as an additional exogenous regressor. The solid line is the median of 10,000 draws from the posterior distribution. Hyperparameters were optimized. The shaded area comprises the 68\% credible region. Lags are those reported in Table \ref{tab:NIF_FEVD} under the gray shaded rows.
\end{figure}

To  open the black box of those rather abstract measurements, we report in Figure \ref{fig:Replication_IRFs} impulse response functions (IRF) for the bivariate models of CO$_2$-GMTA and total forcing-GMTA. Since \cite{sims1980}, the dominant approach for studying the properties of the VAR around its deterministic path has been IRFs to \textit{structural shocks}. Their dynamic effect can be analyzed as that of a randomly assigned treatment because those have been transformed to be uncorrelated, which provides the "keeping everything else constant" interpretation. 

The IRF of a variable $i$ to a one standard deviation shock of $\varepsilon_{j,t}$ is defined as
\begin{align} \label{equ:IRF}
IRF(j\rightarrow i, h) =E(y_{i,t+h}|\boldsymbol{y_{t}},\varepsilon_{t,j}=\sigma_{\varepsilon_{j}})-E(y_{i,t+h}|\boldsymbol{y_{t}},\varepsilon_{t,j}=0).
\end{align}
Thus, it is the expected difference, $h$ months after "impact", between a bivariate system that responded to an unexpected CO$_2$ increase, and the same system where no such increase occurred. In a linear VAR, the above takes a closed-form solution in terms of the matrices from \eqref{struct_VAR}. Models are now estimated with Bayesian methods, optimizing the hyperparameters of a standard Minnesota prior, and choosing the number of lags as reported in the gray-shaded rows of Table \ref{tab:NIF_FEVD}. The primary motivation is to obtain valid inference even in the presence of nonstationarity. Point estimates are nearly identical to that of OLS. For details on such choices in the context of climate data and a more thorough (yet introductory) treatment of IRFs, see \cite{VARCTIC}.

\begin{figure}[h!]
	\caption{IRFs: Annual Emissions and Global Temperature}
	\begin{center}
	
	\begin{subfigure}[t]{0.5\textwidth}
		\includegraphics[trim={0mm 0mm 0mm 10mm},clip,width=\textwidth]{{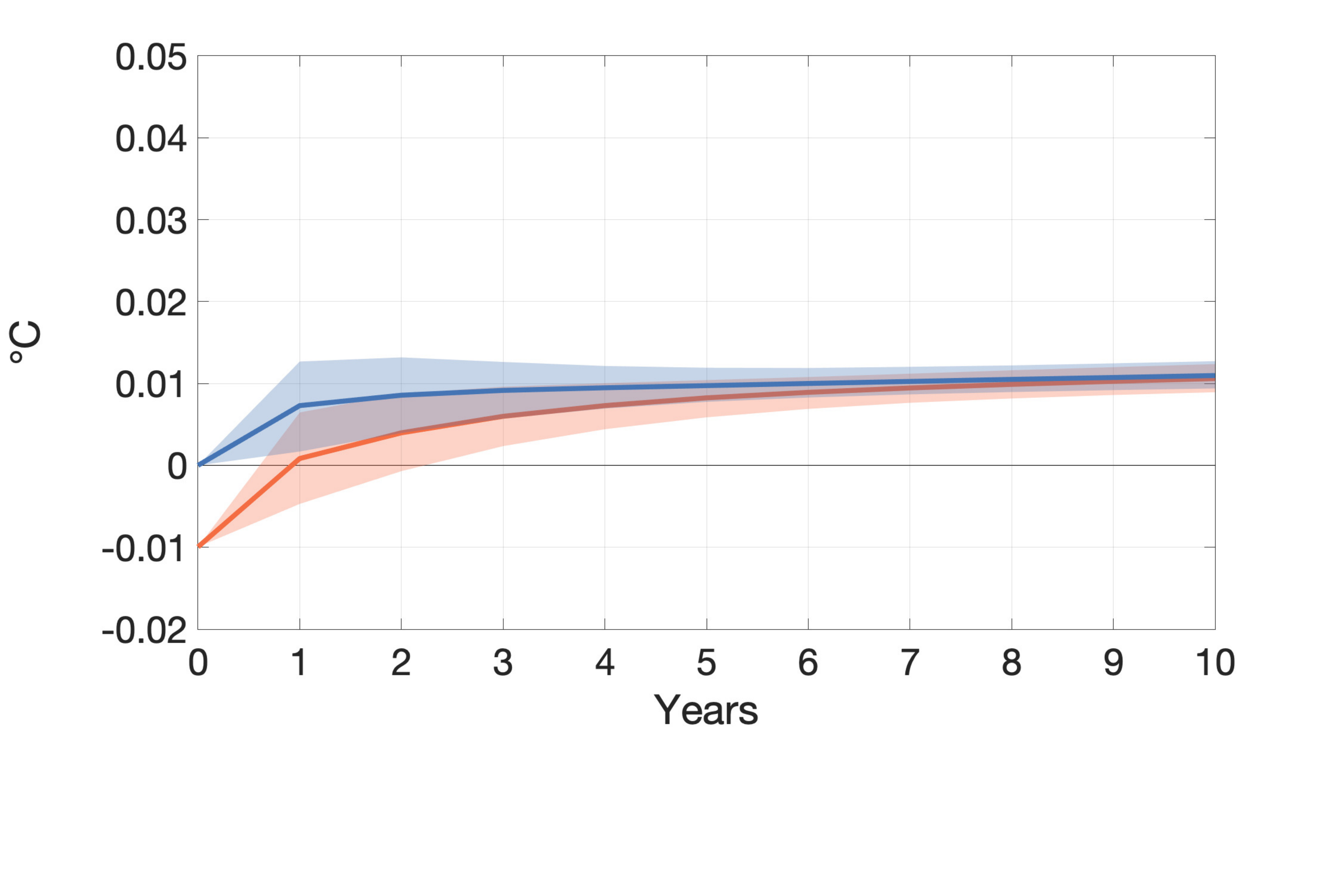}} 
		\vskip -25pt
		\caption{CO$_2$ (Mt/yr) $\rightarrow$  GMTA}
	\end{subfigure}%
	\begin{subfigure}[t]{0.5\textwidth}
		\includegraphics[trim={0mm 0mm 0mm 10mm},clip,width=\textwidth]{{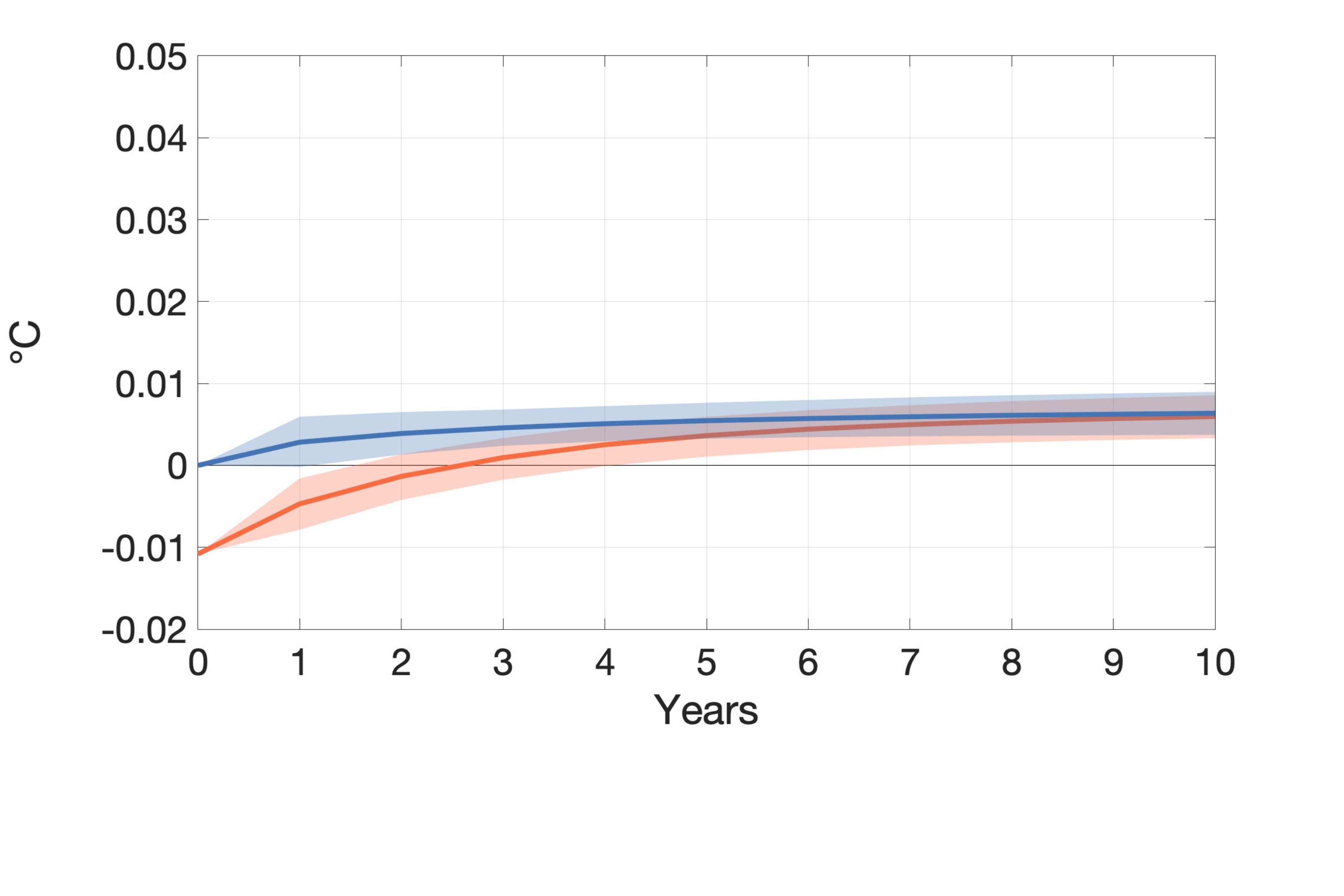}} 
		\vskip -25pt
		\caption{CO$_2$ (Mt/yr) $\rightarrow$  GMTA (including trend)}
	\end{subfigure}\ \\
	
		\vskip 10pt
	
	\begin{subfigure}[t]{0.5\textwidth}
		\includegraphics[trim={0mm 0mm 0mm 10mm},clip,width=\textwidth]{{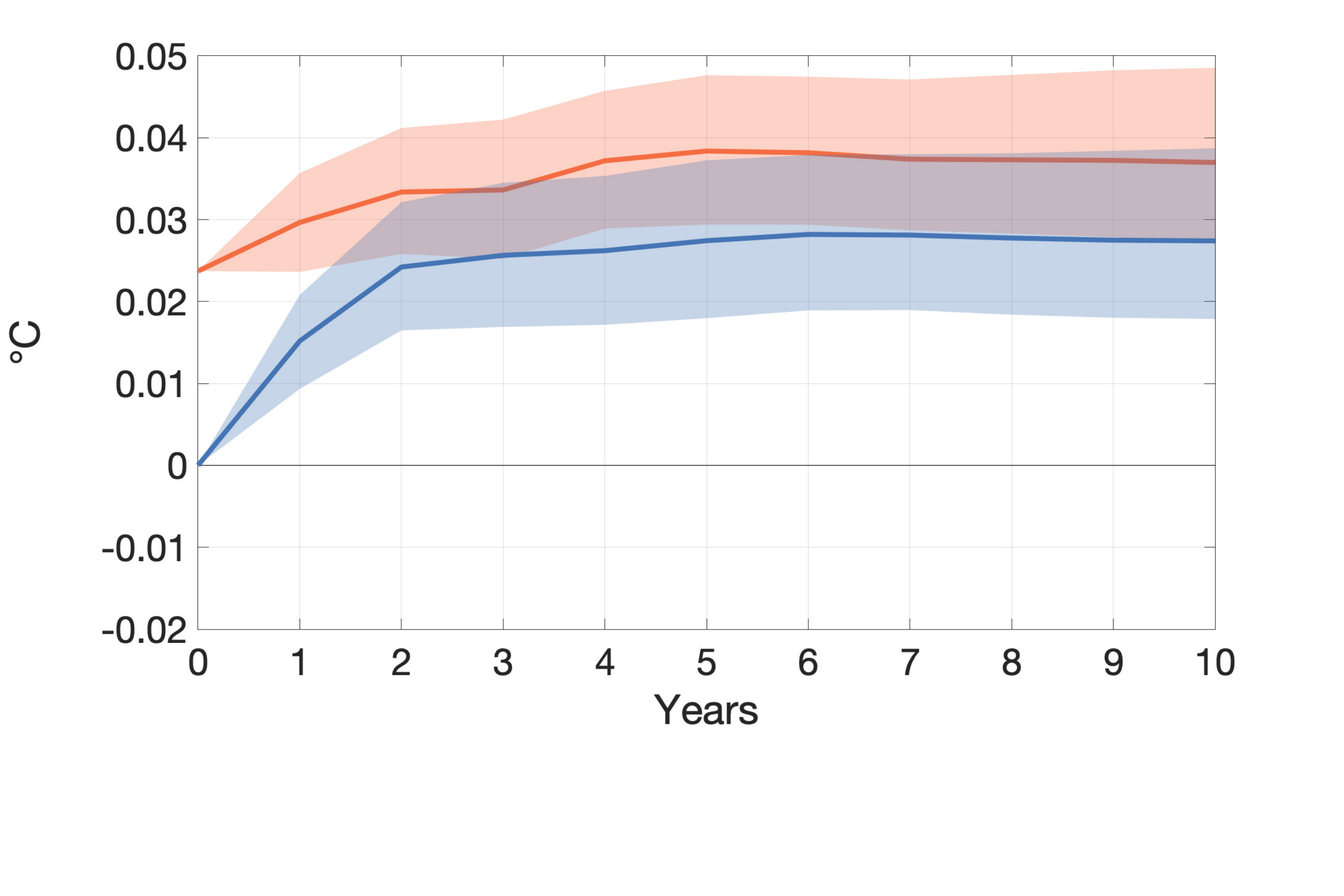}} 
		\vskip -25pt
		\caption{Total Forcing $\rightarrow$  GMTA}
	\end{subfigure}%
	\begin{subfigure}[t]{0.5\textwidth}
		\includegraphics[trim={0mm 0mm 0mm 10mm},clip,width=\textwidth]{{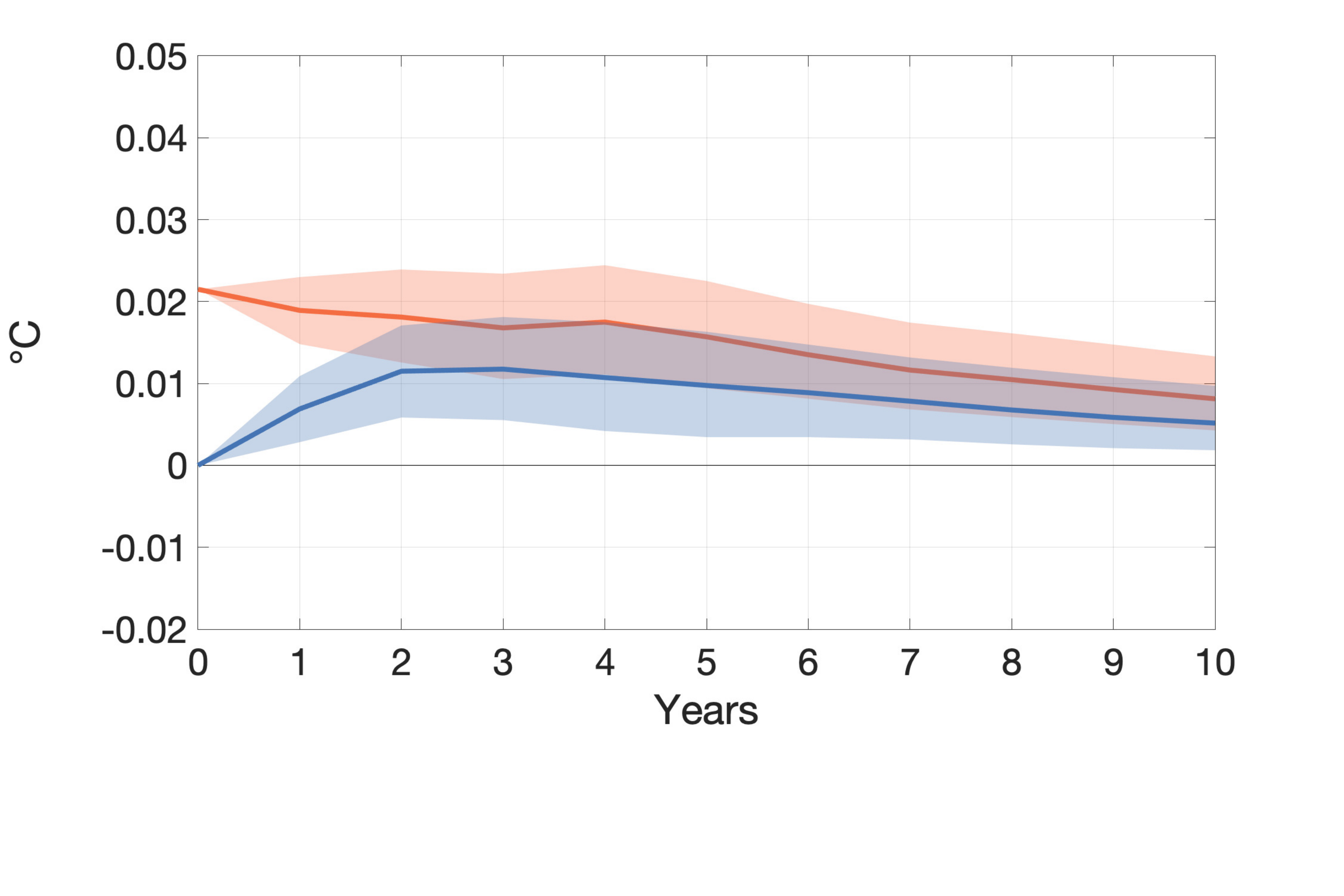}} 
		\vskip -25pt
		\caption{Total Forcing $\rightarrow$ GMTA, (including trend)}
	\end{subfigure}\ \\
		
	\begin{subfigure}[t]{\textwidth}
	\centering
	\vspace*{0.15cm}	
		\includegraphics[trim={0mm 0mm 0mm 0mm},clip,scale=0.3]{{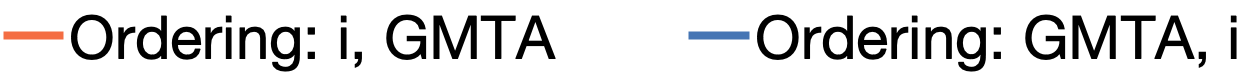}} \vskip 5pt
	\end{subfigure}\\

		\end{center}
	\label{fig:AnnualEmissions_IRFs}
	\scriptsize Notes: We show impulse response functions from bivariate Vector Autoregressions of \emph{GMTA} and annual CO$_2$ emissions. The right column includes a time trend as an additional exogenous regressor. The solid line is the median of 10,000 draws from the posterior distribution. Hyperparameters were optimized. The shaded area comprises the 68\% credible region. Lags are those reported in Table \ref{tab:NIF_FEVD} under the gray shaded rows.
\end{figure}

In Figure \ref{fig:AnnualEmissions_IRFs} we show the effect of an unexpected increase in \textit{annual} emissions on GMTA, while Figure \ref{fig:Replication_IRFs} shows the response of GMTA radiative forcing generated by an unexpected rise in \textit{cumulative} emissions.

Qualitatively, the impact of annual emissions and cumulative CO$_2$-induced forcing \textit{shocks} on global temperature is vastly similar. In accord with findings in \cite{VARCTIC} for the effect of CO$_2$ on Arctic sea ice extent, the impact of total forcing and CO$_2$ shocks is highly durable. This time, it is on GMTA rather than sea ice extent. As reflected in Figures \ref{fig:AnnualEmissions_IRFs} and \ref{fig:Replication_IRFs}, this result is qualitatively independent of the ordering choice. In both cases (and for both forcing variables), the effect of forcing takes about two years to completely settle in. However, it is clear that the reported short-run impact strongly depends on the identification assumptions, which SMCGL completely abstract from. 

Whether the slightly negative short-run response of GMTA in the first panel of Figure \ref{fig:AnnualEmissions_IRFs} favors the ordering $\left\lbrace \text{\textit{GMTA}}, \,\text{\textit{CO$_2$}} \right\rbrace$ over $\left\lbrace \text{\textit{CO$_2$}}, \, \text{\textit{GMTA}} \right\rbrace$ is debatable: \cite{ForsterEtAl2020} find the reduction in global CO$_2$ emissions during the COVID-19 pandemic to have resulted in a short-run rise of global temperature. The key mechanism is a decline in the cooling-effect of aerosols as a result of less SO$_2$ emissions. The authors project a rise in global temperature over the first 24 months following the pandemic-induced reduction in global nitrogen oxide (NO$_x$) emissions.\footnote{Especially NO$_2$ is found to be well-correlated with CO$_2$ emissions \citep{ForsterEtAl2020}.}

Lastly, a robustness check. Our variables are clearly nonstationary. While this does not pose a problem for Bayesian inference, it is nevertheless natural to wonder if results would be significantly altered by including a time trend. Accordingly, the second column reports the same IRFs, but for VAR specifications augmented with trends. Those show that adding such an exogenous explanatory variable does not change the dynamics of a GMTA response to an unexpected shock to CO$_2$. The addition of a trend to the bivariate model of GMTA and total forcing, allows GMTA to slowly revert to a lower impact -- which is nevertheless highly persistent.

\subsection{Are VAR Estimates Quantitatively Reasonable?} \label{sec:empirics_TCR}
The transient climate response (TCR) is a frequently used metric to measure the impact of rising atmospheric CO$_2$ concentration on temperature. It is not only an indication of the trajectory of ongoing climate change, but also serves as a benchmark to evaluate the results of climate model projections \citep{PhillipsLeirvikStorelvmo2020}. The TCR is defined as the increase in temperature, between $h_0$ and $h_T$, under the assumption that CO$_2$ increases annually by 1\%. $h_T$ is defined as that point in time when -- due to the steady annual increase of 1\% -- CO$_2$ concentration is twice as high as at date $h_0$ \citep{Pretis2020,MontamatStock2020}. 
Such a doubling of CO$_2$ would occur approximately after 70 years \citep{OttoEtAl2013}. Following the transformation of \textit{ppm} to \textit{W/m$^\text{2}$} as suggested by \cite{MyhreEtAl2013} and \cite{MontamatStock2020}, a doubling of CO$_2$ under an annual increase in concentration of 1\% would generate a radiative forcing of $5.35 \times \text{ln}\left(2\right) \approx 3.7 \text{ W/m}^2$.\footnote{{An annual increase of 1\% in atmospheric CO$_2$ concentration results in a doubling of CO$_2$ after approximately 70 years, which is described more formally as: $h \times ln\left(\frac{1.01}{1}\right) = ln\left(2\right) \; ,$ for $h \approx 70$ \citep{MontamatStock2020}. }}

Typical estimates for TCR fall within a range of 1$^{\circ}$C-2.5$^{\circ}$C with a 66\% probability, as summarized in the \textit{IPCC 5th Assessment Report} \citep{BindoffEtAl2013}. More recent estimates are well aligned with this range. \cite{BrunsEtAl2020} reports a point estimate of TCR ranging from 1.17$^{\circ}$C to 1.85$^{\circ}$C, depending on the type of data and model specification. \cite{Pretis2020} embeds a two-component energy balance model into a cointegrated vector autoregressive model. His estimates vary across model specification and range from 1.24$^{\circ}$C to 1.38$^{\circ}$C. \cite{PhillipsLeirvikStorelvmo2020} report a global transient climate sensitivity of 2.05$^{\circ}$C. The IV regression in \cite{MontamatStock2020} allows for a differentiation of TCR measurements across different horizons, normalized to giving 70-year-horizon estimates. Their point estimates range in the neighborhood of 1.5$^{\circ}$C within a 95\% confidence interval of roughly 0.9$^{\circ}$C to 2.1$^{\circ}$C.

Despite being much more simplistic than the models of the aforementioned works on TCR, the bivariate VARs do also allow to estimate the impact of a doubling of CO$_2$ on temperature. Here we make use of the concept of IRFs, as presented in Equation \eqref{equ:IRF}. In particular, we estimate the impact on temperature when $RF^{\text{CO}_2}$ increases by one standard deviation of its reduced-form residuals, $\sigma_{\varepsilon_{j}}$, where $j = RF^{\text{CO}_2}$, from a bivariate VAR of $RF^{\text{CO}_2}$ and GMTA.
Recalling the definition of TCR, a doubling of CO$_2$ concentration, which is achieved by an annual increase of 1\% in atmospheric CO$_2$ concentration, generates an additional radiative forcing of approximately $5.35 \times \text{ln}\left(2\right) \approx 3.7 \text{ W/m}^2$. In our case, the shock $\sigma_{\varepsilon_{j}}$ to CO$_2$ is a one-time event at horizon $h = 0$, but its effects are distributed over horizons $h = 1, 2, 3, ..., H$. This allows us to measure the \textit{cumulative} increase, $\Xi_i$, in $i \in \left\lbrace RF^{\text{CO}_2}, GMTA \right\rbrace$ generated by $\sigma_{\varepsilon_{j}}$, where $j = RF^{\text{CO}_2}$, at any horizon $h$:
\[
\Xi_{j,h} = \sum^h_{s=0} IRF(j\rightarrow i, s) \enskip ,
\]
where $IRF(j\rightarrow i, s)$ is defined as in Equation \eqref{equ:IRF}. Adapting the formula of \cite{OttoEtAl2013}, we estimate TCR$_h$ as the increase in GMTA at horizon $h$ as follows:

\begin{align}
TCR_h = \Xi_{GMTA,h} \times \frac{5.35 \times \text{ln}\left(2\right)}{ \Xi_{RF^{\text{CO}_2},h}} \enskip ,
\end{align}

\noindent where $\Xi_{GMTA,h}$ is the cumulative increase in global temperature at horizon $h$, resulting from the shock $\sigma_{\varepsilon_{j}}$, where $j = RF^{\text{CO}_2}$, at horizon $h = 0$. Likewise, $\Xi_{RF^{\text{CO}_2},h}$ is the cumulative increase in radiative forcing of CO$_2$ at horizon $h$, resulting from the shock $\sigma_{\varepsilon_{j}}$, where $j = RF^{\text{CO}_2}$, at horizon $h = 0$.

In Table \ref{tab:TCR} we present median point estimates of TCR$_h$ for $h=20$ and $h=70$ (as in \cite{MontamatStock2020}) from bivariate VARs of $RF^{\text{CO}_2}$ and GMTA -- with and without an exogenous time trend. Thus, Table \ref{tab:TCR} reports the TCR corresponding to the model specifications in Figure \ref{fig:Replication_IRFs} (a) and (b). That is, we use Bayesian estimation techniques and deploy a Minnesota prior on our parameter estimates. Our VAR has four lags and the estimation is based on annual observations between 1850 and 2005.

\begin{table} [h!]
\renewcommand{\arraystretch}{1.2}
	\caption{Transient Climate Response}
	\vspace*{-0.5cm}
	\label{tab:TCR}
	\begin{center}
		{ \footnotesize
		\begin{adjustbox}{max width=\textwidth}
			\begin{tabular}{l c cc  c cc}
				\toprule \toprule \addlinespace[5pt]
				
 \multicolumn{1}{c}{\multirow{2}{*}{Ordering}} 	& &   \multicolumn{2}{c}{\multirow{1}{*}{Without Trend}} & &  \multicolumn{2}{c}{\multirow{1}{*}{With Trend}}\\	
				\cmidrule(lr){3-4} \cmidrule(lr){6-7} \addlinespace[1pt]
				
	& &    \multicolumn{1}{c}{\multirow{1}{*}{TCR$_{20}$}} &  \multicolumn{1}{c}{\multirow{1}{*}{TCR$_{70}$}} & &   \multicolumn{1}{c}{\multirow{1}{*}{TCR$_{20}$}} &  \multicolumn{1}{c}{\multirow{1}{*}{TCR$_{70}$}} \\			
 \cmidrule(lr){1-1} \cmidrule(lr){3-3} \cmidrule(lr){4-4}  \cmidrule(lr){6-6} \cmidrule(lr){7-7}  \addlinespace[5pt]
				 
 \multicolumn{1}{c}{CO$_2$, GMTA} & & 1.99$^{\circ}$C & 2.06$^{\circ}$C & & 2.17$^{\circ}$C & 2.58$^{\circ}$C \\ 	 \addlinespace[5pt]		
 
  \multicolumn{1}{c}{GMTA, CO$_2$} & & 0.57$^{\circ}$C & 1.82$^{\circ}$C & & 0.85$^{\circ}$C & 2.39$^{\circ}$C \\ 	 \addlinespace[1pt]			

\bottomrule	 \bottomrule
			\end{tabular}
			\end{adjustbox}
		}
	\end{center}
\end{table}

The main message of Table \ref{tab:TCR} is twofold: first, the ordering of the variables heavily influences the final results for $\text{TCR}_{20}$, demonstrating the importance of respecting the possibility of cross-correlated residuals. {Mechanically, the discordance brought up by the ordering choice vanishes at much longer horizons, and, as a result, $\text{TCR}_{70}$ estimates are largely similar. However, at that horizon, it is the choice of whether or not to include a trend that can alter results significantly.} Second, even though the TCR$_h$ point estimates of the trend models are rather located at the upper bound of the IPCC range of 1$^{\circ}$C-2.5$^{\circ}$C, a simplistic bivariate VAR model including a constant and a time trend as additional exogenous regressors, is capable of providing a reasonable approximation of the rise in global mean temperature, triggered by a doubling of atmospheric CO$_2$ concentration. 

TCR estimates can be helpful in choosing which ordering is most plausible. For instance, only by ordering CO$_2$ first do we get $\text{TCR}_{20}$ to fall within the IPCC range. IRFs can also help sort things out. Ordering CO$_2$ second leads to a surprisingly lasting \textit{negative} effect of CO$_2$ shocks on GMTA. An increasingly popular approach to VAR identification in macroeconomics is to use sign restrictions, where implausible IRF draws (based on economic theory) are tossed out \citep{Uhlig2005}. This dispenses the researcher from formulating a likely contentious causal ordering of variables. In a climate application, one could identify the VAR by rejecting specifications generating implausible $IRF(\text{CO$_2$}\rightarrow \text{GMTA}, h)$ or TCRs. Applying this sort of reasoning leads us to favor -- to nobody's surprise -- the specification where simultaneous causality runs from CO$_2$ to GMTA. However, it is important to stress that this choice is obtained from prior knowledge on what is deemed reasonable and what is not, rather than our two time series.

\section{Conclusion}\label{sec:con}
This note is a cautionary tale about how seemingly innocuous simplifying assumptions can go wrong -- especially when they are formulated without consulting the data. IFs, as proposed by SMCGL, is a concept that hinges on the assumption of zero correlation between the residuals of a bivariate VAR(1) process. In discrete time, especially with observations at lower frequencies, such an assumption is most often not justified. Both stylized simulations and an empirical application in the form of the transient climate response demonstrate that being negligent about cross-correlated residuals can lead to markedly different outcomes. Our results show that already in a bivariate system of CO$_2$ and GMTA, the resulting TCR depends on how one deals with correlated residuals. 

FEVDs provide an alternative to IFs. Albeit being a decisive improvement over IFs, FEVDs are as good as their underlying statistical model. To avoid departing too much from the SMCGL's framework, we only considered bivariate models. Climate systems obviously comprise numerous additional variables. \cite{wilson2010atmospheric} considers four in a VAR setup, but there could be many more. Additionally, the dynamics were assumed to be linear and time invariant, an approximation that should be eventually tested. \cite{estrada2013statistically} trend "breaks" model is one way to do it and they report results pointing in the same direction as ours. However, with machine learning tools becoming increasingly common use, more flexible alternatives could be used to yield further insights. Finally, we considered simplistic identification schemes which implied a causal ordering of variables. There is a plethora of more sophisticated schemes available \citep{kilian2017svar} and those could be used in future work, especially when moving beyond bivariate systems. Another -- simpler -- avenue is the use of data sampled at higher frequencies (like daily) which, by construction, makes the simultaneity problem much less of a Damocles sword.


\clearpage

\bibliographystyle{apalike}
	\bibliography{Liang2} 

\begin{thebibliography}{}

\bibitem[Andrews et~al., 2010]{AndrewsEtAl2010}
Andrews, T., Forster, P., Boucher, O., Bellouin, N., and Jones, A. (2010).
\newblock {Precipitation, Radiative Forcing and Global Temperature Change}.
\newblock {\em Geophysical Research Letters}, 37(14).

\bibitem[Bindoff et~al., 2013]{BindoffEtAl2013}
Bindoff, N., Stott, P., AchutaRao, P., Allen, M., Gillett, N., Gutzler, D.,
  Hansingo, K., Hegerl, G., Hu, Y., Jain, S., Mokhov, I., Overland, J.,
  Perlwitz, J., Sebbari, R., and Zhang, X. (2013).
\newblock {Chapter 10 - Detection and Attribution of Climate Change: From
  Global to Regional}.
\newblock In {\em {Climate Change 2013: The Physical Science Basis. IPCC
  Working Group I Contribution to AR5}}. Cambridge University Press, Cambridge.

\bibitem[Bruns et~al., 2020]{BrunsEtAl2020}
Bruns, S., Csereklyei, Z., and Stern, D. (2020).
\newblock {A Multicointegration Model of Global Climate Change}.
\newblock {\em Journal of Econometrics}, 214(1):175--197.
\newblock Annals Issue: Econometric Models of Climate Change.

\bibitem[Chen et~al., 2011]{chen2011vector}
Chen, G., Glen, D.~R., Saad, Z.~S., Hamilton, J.~P., Thomason, M.~E., Gotlib,
  I.~H., and Cox, R.~W. (2011).
\newblock Vector autoregression, structural equation modeling, and their
  synthesis in neuroimaging data analysis.
\newblock {\em Computers in biology and medicine}, 41(12):1142--1155.

\bibitem[Estrada et~al., 2013]{estrada2013statistically}
Estrada, F., Perron, P., and Mart{\'\i}nez-L{\'o}pez, B. (2013).
\newblock Statistically derived contributions of diverse human influences to
  twentieth-century temperature changes.
\newblock {\em Nature Geoscience}, 6(12):1050--1055.

\bibitem[Etminan et~al., 2016]{EtminanEtAl2016}
Etminan, M., Myhre, G., Highwood, E., and Shine, K. (2016).
\newblock {Radiative Forcing of Carbon Dioxide, Methane, and Nitrous Oxide: A
  significant Revision of the Methane Radiative Forcing}.
\newblock {\em Geophysical Research Letters}, 43(24):12,614--12,623.

\bibitem[Forster et~al., 2020]{ForsterEtAl2020}
Forster, P., Forster, H., Evans, M., Gidden, M., Jones, C., Keller, C.,
  Lamboll, R., Qu{\'e}r{\'e}, C., Rogelj, J., Rosen, D., Schleussner, C.-F.,
  Richardson, T., Smith, C., and Turnock, S. (2020).
\newblock {Current and Future Global Climate Impacts Resulting from COVID-19}.
\newblock {\em Nature Climate Change}, 10(10):913--919.

\bibitem[Goulet~Coulombe and G{\"o}bel, 2021]{VARCTIC}
Goulet~Coulombe, P. and G{\"o}bel, M. (2021).
\newblock Arctic amplification of anthropogenic forcing: A vector
  autoregressive analysis.
\newblock {\em Journal of Climate}, Forthcoming.

\bibitem[Granger, 1969]{granger1969}
Granger, C.~W. (1969).
\newblock Investigating causal relations by econometric models and
  cross-spectral methods.
\newblock {\em Econometrica: Journal of the Econometric Society}, pages
  424--438.

\bibitem[Granville~Tunnicliffe, 2015]{wilson2010atmospheric}
Granville~Tunnicliffe, W. (2015).
\newblock Atmospheric co2 and global temperatures: the strength and nature of
  their dependence.
\newblock {\em Working Paper}.

\bibitem[Hansen et~al., 2006]{Hansen2006}
Hansen, J., Sato, M., Ruedy, R., Lo, K., Lea, D., and Medina-Elizade, M.
  (2006).
\newblock {Global Temperature Change}.
\newblock {\em Proceedings of the National Academy of Sciences},
  103(39):14288--14293.

\bibitem[Kilian and L{\"u}tkepohl, 2017]{kilian2017svar}
Kilian, L. and L{\"u}tkepohl, H. (2017).
\newblock {\em Structural vector autoregressive analysis}.
\newblock Cambridge University Press.

\bibitem[Koutsoyiannis and Kundzewicz, 2020]{koutsoyiannis2020atmospheric}
Koutsoyiannis, D. and Kundzewicz, Z.~W. (2020).
\newblock Atmospheric temperature and co2: Hen-or-egg causality?
\newblock {\em Sci}, 2(4):83.

\bibitem[Li et~al., 2013]{LiEtAl2013}
Li, C., Notz, D., Tietsche, S., and Marotzke, J. (2013).
\newblock {The Transient versus the Equilibrium Response of Sea Ice to Global
  Warming}.
\newblock {\em Journal of Climate}, 26(15):5624--5636.

\bibitem[Liang, 2008]{Liang2008}
Liang, X. (2008).
\newblock {Information Flow within Stochastic Dynamical System}.
\newblock {\em Phys. Rev. E}, 78(5):031113.

\bibitem[Liang, 2014]{Liang2014}
Liang, X. (2014).
\newblock {Unraveling the Cause-Effect Relation between Time Series}.
\newblock {\em Phys. Rev. E}, 90:052150.

\bibitem[Liang, 2015]{Liang2015}
Liang, X. (2015).
\newblock {Normalizing the Causality between Time Series}.
\newblock {\em Phys. Rev. E}, 92:022126.

\bibitem[Liang, 2016]{Liang2016}
Liang, X. (2016).
\newblock {Information Flow and Causality as rigorous Notions ab initio}.
\newblock {\em Phys. Rev. E}, 94:052201.

\bibitem[Marotzke and Forster, 2015]{Marotzke2015}
Marotzke, J. and Forster, P. (2015).
\newblock {Forcing, Feedback and Internal Variability in Global Temperature
  Trends}.
\newblock {\em Nature}, 517(7536):565--570.

\bibitem[Masson-Delmotte et~al., 2018]{ipccGlobWarm2018}
Masson-Delmotte, V., Zhai, P., P\"ortner, H.-O., Roberts, D., Skea, J., Shukla,
  P., Pirani, A., Moufouma-Okia, W., P\'ean, C., Pidcock, R., Connors, S.,
  Matthews, J.B.R.and~Chen, Y., Zhou, X., Gomis, M., Lonnoy, E., Maycock, T.,
  Tignor, M., and Waterfield, T. (2018).
\newblock {IPCC, 2018: Global Warming of 1.5${}^\circ$C. An IPCC Special Report
  on the Impacts of Global Warming of 1.5${}^\circ$C above Pre-Industrial
  Levels and related Global Greenhouse Gas Emission Pathways, in the Context of
  Strengthening the Global Response to the Threat of Climate Change,
  Sustainable Development, and Efforts to Eradicate Poverty}.
\newblock {\em In press}.

\bibitem[Meinshausen et~al., 2017]{Meinshausen2017}
Meinshausen, M., Vogel, E., Nauels, A., Lorbacher, K., Meinshausen, N.,
  Etheridge, D., Fraser, P., Montzka, S., Rayner, P., Trudinger, C., Krummel,
  P., Beyerle, U., Canadell, J., Daniel, J., Enting, I.~G., Law, R., Lunder,
  C., O'Doherty, S., Prinn, R., Reimann, S., Rubino, M., Velders, G., Vollmer,
  M., Wang, R., and Weiss, R. (2017).
\newblock Historical greenhouse gas concentrations for climate modelling
  (cmip6).
\newblock {\em Geoscientific Model Development}, 10(5):2057--2116.

\bibitem[Montamat and Stock, 2020]{MontamatStock2020}
Montamat, G. and Stock, J. (2020).
\newblock Quasi-experimental estimates of the transient climate response using
  observational data.
\newblock {\em Climatic Change}, 160(3):361--371.

\bibitem[Myhre et~al., 2013]{MyhreEtAl2013}
Myhre, G., Shindell, D., Bréon, F.-M., Collins, W., Fuglestvedt, J., Huang,
  J., Koch, D., Lamarque, J.-F., Lee, D., Mendoza, B., Nakajima, T., Robock,
  A., Stephens, G., Takemura, T., and Zhang, H. (2013).
\newblock {Anthropogenic and Natural Radiative Forcing}.
\newblock In Stocker, T., Qin, D., Plattner, G.-K., Tignor, M., Allen, S.~K.,
  Doschung, J., Nauels, A., Xia, Y., Bex, V., and Midgley, P., editors, {\em
  {Climate Change 2013: The Physical Science Basis. Contribution of Working
  Group I to the Fifth Assessment Report of the Intergovernmental Panel on
  Climate Change}}, pages 659--740. Cambridge University Press, Cambridge, UK.

\bibitem[Nordhaus, 2014]{Nord14}
Nordhaus, W. (2014).
\newblock Estimates of the social cost of carbon: Concepts and results from the
  dice-2013r model and alternative approaches.
\newblock {\em Journal of the Association of Environmental and Resource
  Economists}, 1(1/2):273--312.

\bibitem[Notz and Stroeve, 2016]{notz2016observed}
Notz, D. and Stroeve, J. (2016).
\newblock Observed arctic sea-ice loss directly follows anthropogenic co2
  emission.
\newblock {\em Science}, 354(6313):747--750.

\bibitem[Otto et~al., 2013]{OttoEtAl2013}
Otto, A., Otto, F., Boucher, O., Church, J., Hegerl, G., Forster, P., Gillett,
  N.~P., Gregory, J., Johnson, G., Knutti, R., Lewis, N., Lohmann, U.,
  Marotzke, J., Myhre, G., Shindell, D., Stevens, B., and Allen, M. (2013).
\newblock {Energy Budget Constraints on Climate Response}.
\newblock {\em Nature Geoscience}, 6(6):415--416.

\bibitem[Phillips et~al., 2020]{PhillipsLeirvikStorelvmo2020}
Phillips, P., Leirvik, T., and Storelvmo, T. (2020).
\newblock {Econometric Estimates of Earth’s Transient Climate Sensitivity}.
\newblock {\em Journal of Econometrics}, 214(1):6--32.
\newblock Annals Issue: Econometric Models of Climate Change.

\bibitem[Pretis, 2020]{Pretis2020}
Pretis, F. (2020).
\newblock {Econometric Modelling of Climate Systems: The Equivalence of Energy
  Balance Models and Cointegrated Vector Autoregressions}.
\newblock {\em Journal of Econometrics}, 214(1):256--273.
\newblock Annals Issue: Econometric Models of Climate Change.

\bibitem[Shukla et~al., 2019]{ipcc2019}
Shukla, P., Skea, J., Calvo~Buendia, E., Masson-Delmotte, V., Pörtner, H.-O.,
  Roberts, D., Zhai, P., Slade, R., Connors, S., van Diemen, R., Ferrat, M.,
  Haughey, E., Luz, S., Neogi, S., Pathak, M., Petzold, J., Portugal~Pereira,
  J., Vyas, P., Huntley, E., Kissick, K., Belkacemi, M., and Malley, J. (2019).
\newblock {Climate Change and Land: an IPCC Special Report on Climate Change,
  Desertification, Land Degradation, Sustainable Land Management, Food
  Security, and Greenhouse Gas Fluxes in Terrestrial Ecosystems}.
\newblock {\em In press}.

\bibitem[Sims, 1980]{sims1980}
Sims, C.~A. (1980).
\newblock Macroeconomics and reality.
\newblock {\em Econometrica: journal of the Econometric Society}, pages 1--48.

\bibitem[Stips et~al., 2016]{StipsEtAl2016}
Stips, A., Macias, D., Coughlan, C., Garcia-Gorriz, E., and Liang, X. (2016).
\newblock {On the Causal Structure between CO2 and Global Temperature}.
\newblock {\em Scientific Reports}, 6:21691.

\bibitem[Tawia~Hagan et~al., 2019]{tawia2019time}
Tawia~Hagan, D.~F., Wang, G., San~Liang, X., and Dolman, H.~A. (2019).
\newblock A time-varying causality formalism based on the liang--kleeman
  information flow for analyzing directed interactions in nonstationary climate
  systems.
\newblock {\em Journal of Climate}, 32(21):7521--7537.

\bibitem[Tollefson, 2014]{Tollefson14}
Tollefson, J. (2014).
\newblock {Climate Change: The Case of the Missing Heat}.
\newblock {\em Nature}, 505:276--278.

\bibitem[Uhlig, 2005]{Uhlig2005}
Uhlig, H. (2005).
\newblock What are the effects of monetary policy on output? results from an
  agnostic identification procedure.
\newblock {\em Journal of Monetary Economics}, 52(2):381 -- 419.

\end{thebibliography}
  
  \clearpage

\appendix
\newcounter{saveeqn}
\setcounter{saveeqn}{\value{section}}
\renewcommand{\theequation}{\mbox{\Alph{saveeqn}.\arabic{equation}}} \setcounter{saveeqn}{1}
\setcounter{equation}{0}


\section{Data Sources}

For the empirical part of Section \ref{sec:empirics}, we refer to the variables presented in  Table 1 of SMCGL. Due to the sparse description of the data sources, our data set reduces to the annual global mean surface temperature anomalies (GMTA) and seven other representatives of radiative forcing. For five variables we could match the reported correlation coefficients as well as the IF values of SMCGL, Table \ref{tab:NIF_FEVD}. For \emph{total forcing} and \emph{solar} these measures do not coincide. For the analysis in Section \ref{sec:empirics_IRFs} and  \ref{sec:empirics_TCR}, we use the CO$_2$ series based on \cite{Meinshausen2017}. We transform \textit{ppm} into W/m$^\text{2}$ as described in Section \ref{sec:empirics}.

\begin{table}[H]
  \begin{threeparttable}
  \centering
  \scriptsize
  \caption{List of Variables}\label{tab:ListofAbb}
  \setlength{\tabcolsep}{0.1em}
    \begin{tabularx}{\textwidth}{sXs}
    \toprule \toprule
    \textbf{Abbreviation} & \textbf{Description} & \textbf{Data Source} \\
    \midrule
 \rowcolor{gray!15}
Total Forcing & annual; 1850-2005 & 
\href{https://climexp.knmi.nl/getindices.cgi?WMO=LeedsData/Total_ERF}{KNMI Climate Explorer} \\ 
   
Anthropogenic & annual; 1850-2005 & 
\href{https://climexp.knmi.nl/getindices.cgi?WMO=LeedsData/Anthropogenic_total_ERF}{KNMI Climate Explorer} \\ 
 
   \rowcolor{gray!15}   
 CO$_2$-ERF (W/m$^\text{2}$)  & annual; 1850-2005 & 
 \href{https://climexp.knmi.nl/getindices.cgi?WMO=LeedsData/CO2_ERF}{KNMI Climate Explorer} \\
      
Aerosol & annual; 1850-2005 & 
\href{https://climexp.knmi.nl/getindices.cgi?WMO=LeedsData/total_aerosol_ERF}{KNMI Climate Explorer}\\

   \rowcolor{gray!15}  
Solar & annual; 1850-2005 & 
\href{https://climexp.knmi.nl/getindices.cgi?WMO=LeedsData/Solar_ERF}{KNMI Climate Explorer} \\

Volcanic & annual; 1850-2005 & 
\href{https://climexp.knmi.nl/getindices.cgi?WMO=LeedsData/Volcanic_ERF}{KNMI Climate Explorer}  \\

   \rowcolor{gray!15}  
PDO & annual averages of monthly observations; 1900-2005 & 
{\href{https://climexp.knmi.nl/getindices.cgi?WMO=UWData/pdo&STATION=PDO&TYPE=i}{KNMI Climate Explorer}} \\

GMTA & annual; global; 1900-2005 & 
\href{https://crudata.uea.ac.uk/cru/data/temperature/}{HadCRUT4}\\

   \rowcolor{gray!15}   
CO$_2$ (Million Tonnes / Year) & annual; global production-based emissions; 1850-2005 & 
 \href{https://github.com/owid/co2-data}{Our World in Data} -- not in use\\
 
 CO$_2$ (ppm) & \makecell[l]{ annual; global; 1850-2014 -- based on \cite{Meinshausen2017} \\ annual; global; 2015-2017 -- based on NOAA-ESRL} & \makecell[l]{ 1850-2014: 
 \href{ftp://data.iac.ethz.ch/CMIP6/input4MIPs/UoM/GHGConc/CMIP/yr/atmos/UoM-CMIP-1-1-0/GHGConc/gr3-GMNHSH/v20160701/mole_fraction_of_carbon_dioxide_in_air_input4MIPs_GHGConcentrations_CMIP_UoM-CMIP-1-1-0_gr3-GMNHSH_0000-2014.csv}{IAC ETH Z\"urich} \\ 2015-2017: \href{https://www.esrl.noaa.gov/gmd/webdata/ccgg/trends/co2/co2_annmean_gl.txt}{NOAA-ESRL}}\\

\bottomrule \bottomrule
    \end{tabularx}
  \end{threeparttable}
  \end{table}

\clearpage

\section{Additional Simulation Results} \label{sec:add_simulations}

In Figure \ref{fig:DIFFs_TauPlot_matching_h2} below, we compare NIF and FEVD$_{\boldsymbol{h=2}}$. One might be concerned that results in Figure \ref{fig:DIFFs_TauPlot_matching} suffer from a horizon mismatch between NIF and FEVD$_{h=10}$.  Results suggest that reducing $h$ worsens NIFs' problems by shrinking the "safe" regions. This is intuitive, the effect of assumptions on simultaneous relationships gets milder as we get further from $h=0$. 

\begin{figure}[h!]
	\begin{center}
	\begin{subfigure}[t]{0.5\textwidth}
		\includegraphics[trim={0mm 0mm 0mm 10mm},clip,width=\textwidth]{{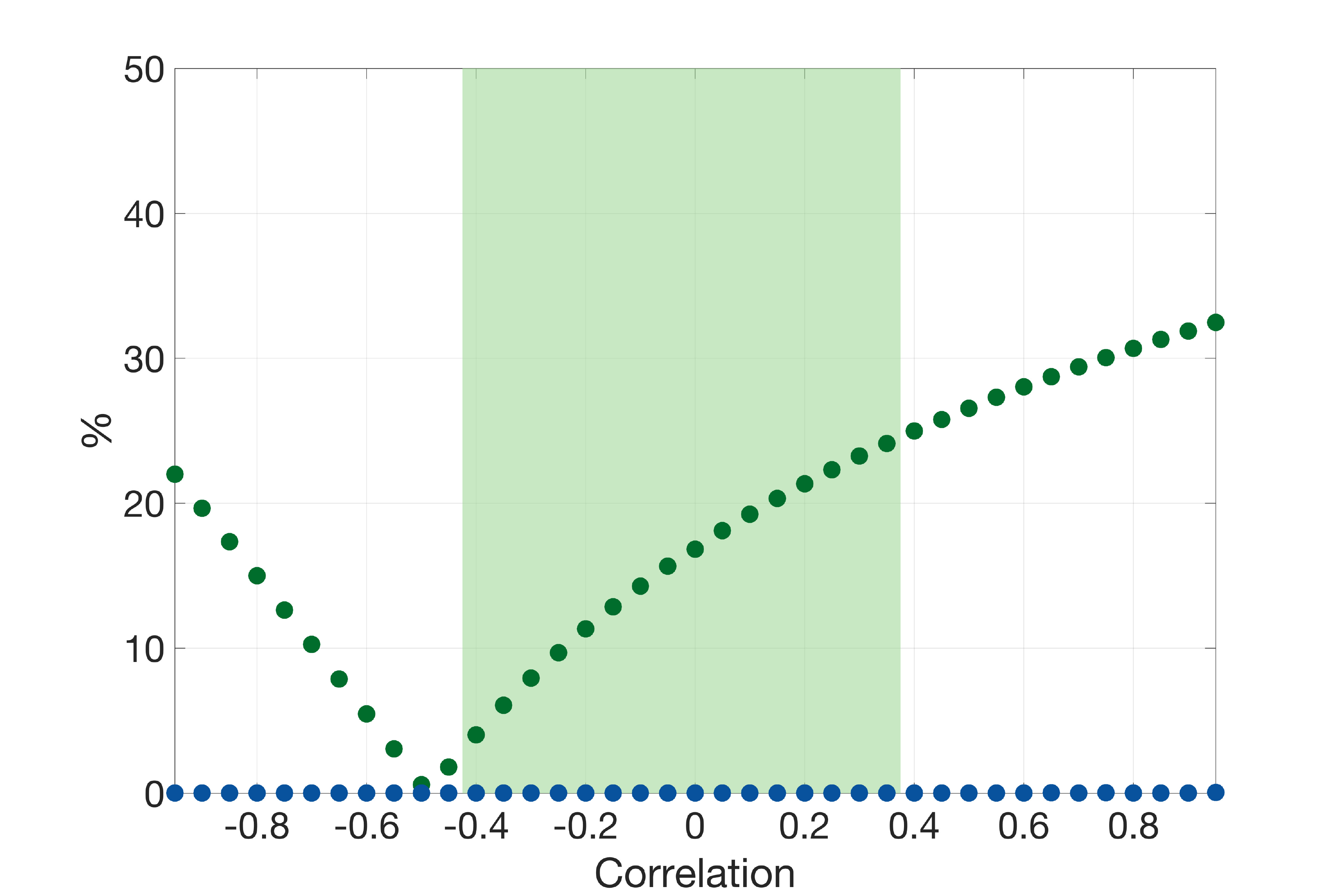}} 
		\caption{DGP(1)} 
	\end{subfigure}%
	\begin{subfigure}[t]{0.5\textwidth}
		\includegraphics[trim={0mm 0mm 0mm 10mm},clip,width=\textwidth]{{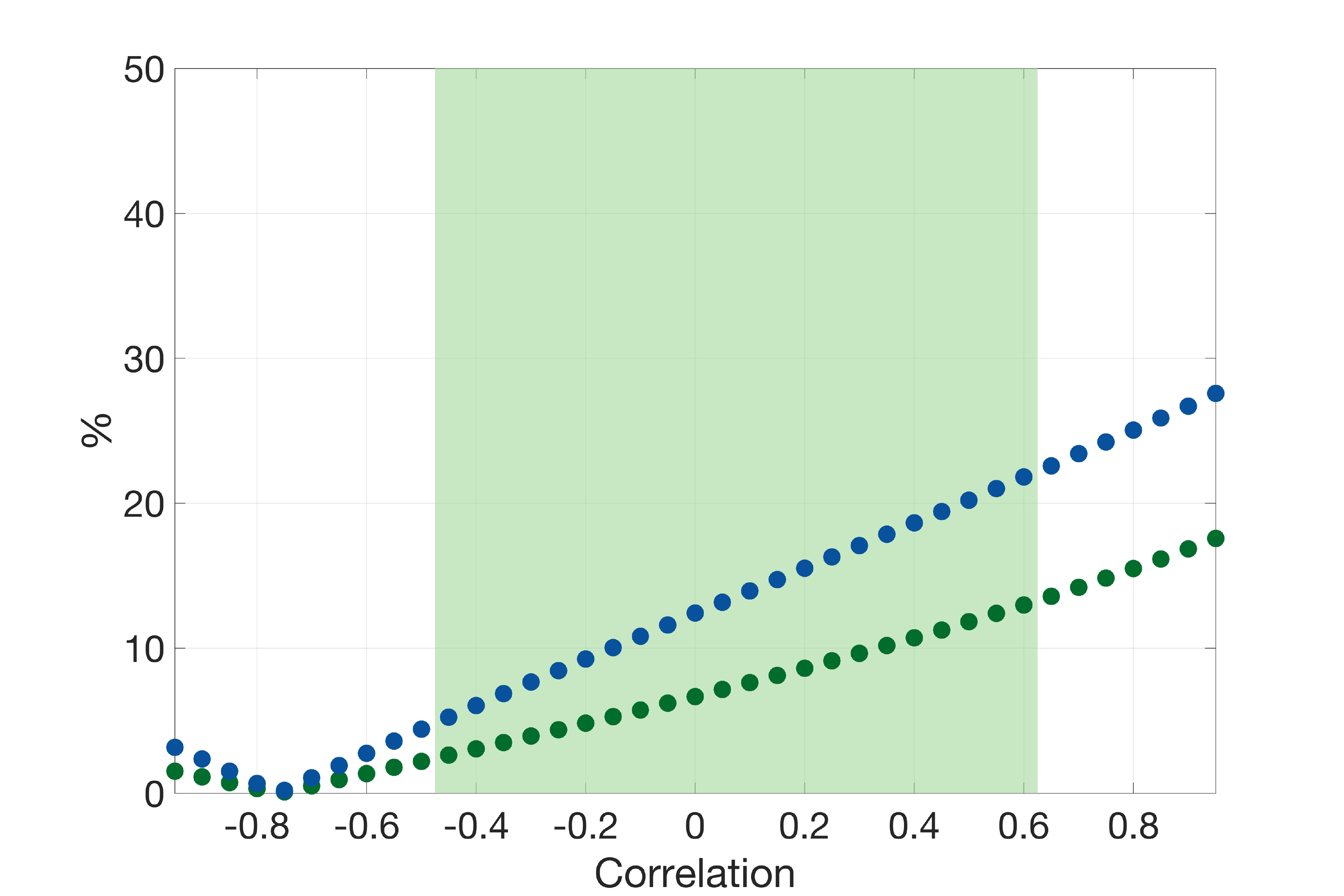}} 
		\caption{DGP(2)} 
	\end{subfigure}
\vskip 10pt
	\begin{subfigure}[t]{0.5\textwidth}
		\includegraphics[trim={0mm 0mm 0mm 10mm},clip,width=\textwidth]{{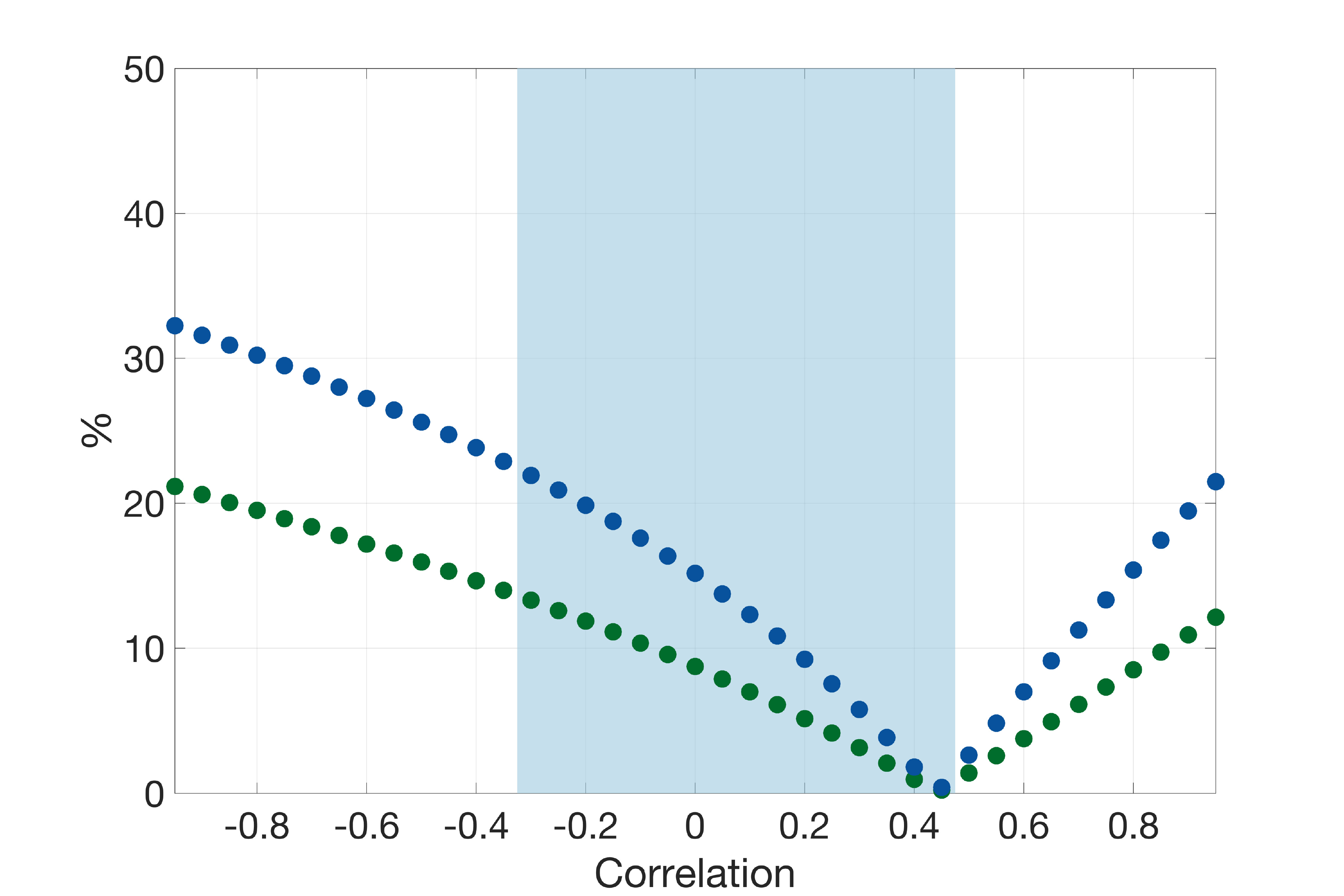}} 
		\caption{DGP(3)} 
	\end{subfigure}%
	\begin{subfigure}[t]{0.5\textwidth}
		\includegraphics[trim={0mm 0mm 0mm 10mm},clip,width=\textwidth]{{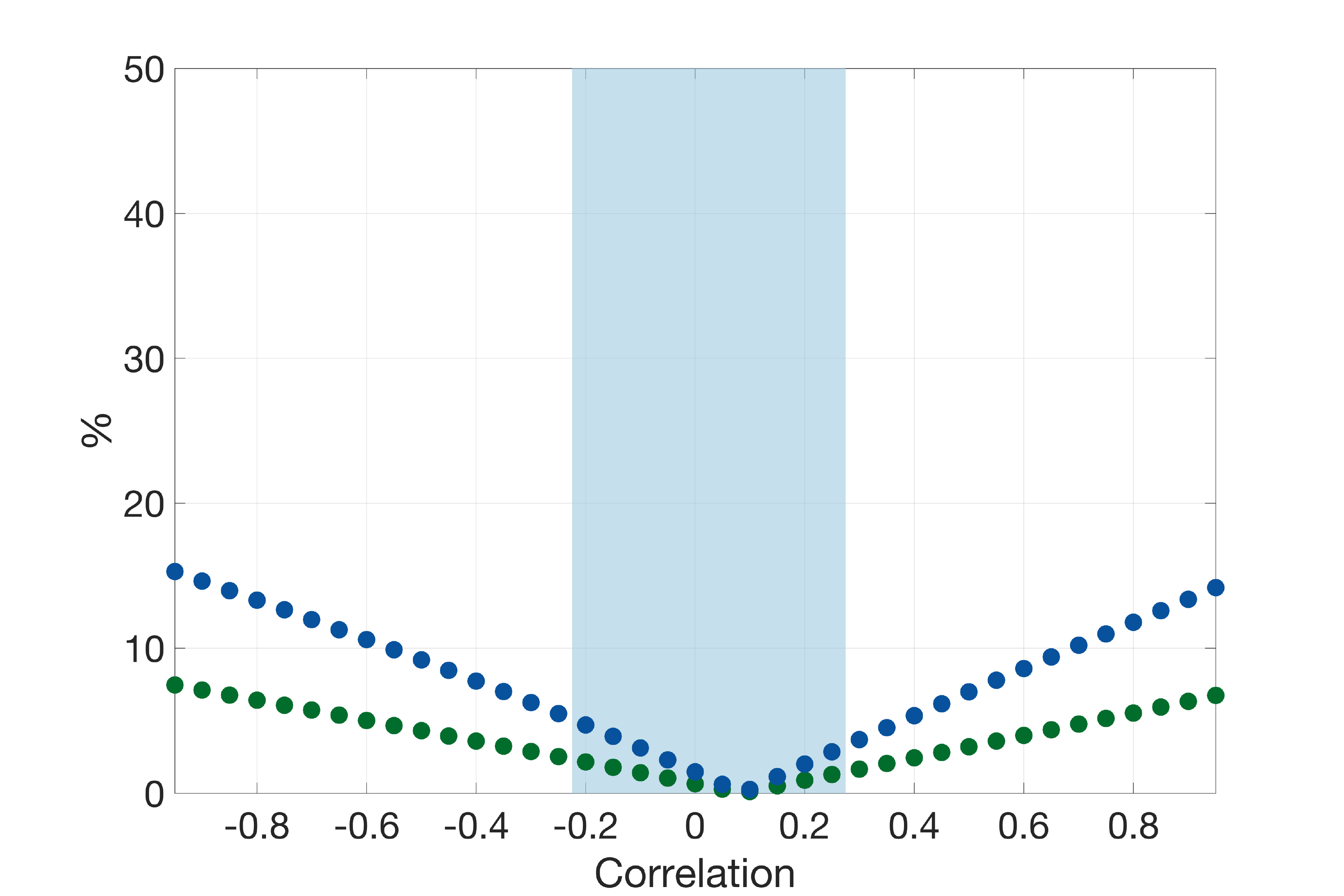}} 
		\caption{DGP(4)}
	\end{subfigure}
	\begin{subfigure}[t]{\textwidth}
	\centering
	\vspace*{0.5cm}	
	\includegraphics[trim={0mm 0mm 0mm 0mm},clip,scale=0.4]{{Figures/TauPlot_FEVD_matchingLegend.png}} \vskip 5pt
	\end{subfigure}
	\vskip 5pt
	\end{center}
	\noindent \scriptsize \textbf{Notes}: Color specification: the \emph{light blue} area shows the region, in which FEVDs suggest $\Upsilon^{i,j}_{1 \rightarrow 2}$ $>$ $\Upsilon^{i,j}_{2 \rightarrow 1}$, regardless of the ordering. Similarly, the \emph{light green} area shows the region, in which FEVDs suggest $\Upsilon^{i,j}_{2 \rightarrow 1}$ $>$ $\Upsilon^{i,j}_{1 \rightarrow 2}$ -- independent of the ordering. The \emph{white/non-colored} areas shows those regions, for which the ordering of $X1$ and $X2$ does matter. An unambiguous determination of $\Upsilon^{i,j}_{1 \rightarrow 2}$ $>$ $\Upsilon^{i,j}_{2 \rightarrow 1}$, respectively $\Upsilon^{i,j}_{2 \rightarrow 1}$ $>$ $\Upsilon^{i,j}_{1 \rightarrow 2}$, is not possible.
	\vskip 5pt
		\caption{Ranking of NIFs ($\tau_{i \rightarrow j}$) vs Ranking of FEVDs ($\Upsilon^{i,j}_{i \rightarrow j}$) \\ Different Levels of Correlation; Horizon $h = 2$}
	\label{fig:DIFFs_TauPlot_matching_h2}
\end{figure}

\clearpage

\section{Additional Empirical Results}

\begin{table} [h!]
\renewcommand{\arraystretch}{1.2}
	\caption{Empirical Results for the Bivariate Relationship \\ Between Various Forcings and GMTA \\ {\footnotesize Sample Period: 1850-\textbf{2017}}}
	\vspace*{-0.5cm}
	\label{tab:NIF_FEVD_18502017}
	\begin{center}
		{ \scriptsize
		\begin{adjustbox}{max width=\textwidth}
			\begin{tabular}{l c cc  c ccccc}
				\toprule \toprule \addlinespace[5pt]
				
				&   \multicolumn{1}{c|}{\multirow{3}{*}{\textbf{Correlation}}} & \multicolumn{2}{c}{\multirow{2}{*}{\textbf{\makecell{ Normalized IF \\ [3pt] (IF $\times$ 100) }}}}& \multicolumn{6}{|c}{\textbf{FEVD}} \\	
				\cmidrule(lr){5-10}
				
				&   \multicolumn{1}{c|}{}                &  & & \multicolumn{1}{|c}{\multirow{2}{*}{\textbf{\makecell{Lags \\ [3pt] ($P$)} }}} & \multirow{2}{*}{\textbf{\makecell{Correlation \\ of \\ Residuals ($\rho_u$)}}} & \multicolumn{2}{c}{\textbf{Ordering:} $i$, GMTA} & \multicolumn{2}{c}{\textbf{Ordering:} GMTA, $i$} \\
				\cmidrule(lr){3-4} \cmidrule(lr){7-8} \cmidrule(lr){9-10}

				&	 &   \multicolumn{1}{|c}{$i$ $\rightarrow$ GMTA} & \multicolumn{1}{c|}{GMTA $\rightarrow$ $i$} & & &
				${i \rightarrow \text{GMTA}}$ & ${\text{GMTA} \rightarrow i}$ &
				${i \rightarrow \text{GMTA}}$ & ${\text{GMTA} \rightarrow i}$ \\  [3pt]
				
\cmidrule(lr){1-1} 	\cmidrule(lr){2-2} \cmidrule(lr){3-3} \cmidrule(lr){4-4}  \cmidrule(lr){5-5} \cmidrule(lr){6-6}  \cmidrule(lr){7-7} \cmidrule(lr){8-8}  \cmidrule(lr){9-9} \cmidrule(lr){10-10}  \addlinespace [5pt]

\multicolumn{1}{l|}{\multirow{2}{*}{Total Forcing}} 
& \multirow{2}{*}{0.82}  & \textbf{36.8} & \multicolumn{1}{c|}{29.5} & 

  \cellcolor{gray!15} 4 &  \cellcolor{gray!15} 0.23*** &
\cellcolor{gray!15} \multirow{1}{*}{\textbf{48.0}} & \cellcolor{gray!15} \multirow{1}{*}{15.6}  & \cellcolor{gray!15} \multirow{1}{*}{28.3}  & \cellcolor{gray!15} \multirow{1}{*}{\textbf{29.3}} \\ 

\multicolumn{1}{l|}{}&   & (17.2) & \multicolumn{1}{c|}{(15.3)} & 
  1 & 0.29*** &
 \multirow{1}{*}{\textbf{55.7}} & \multirow{1}{*}{14.4}  & \multirow{1}{*}{30.0} & \multirow{1}{*}{\textbf{36.6}} 
 \\ \addlinespace[5pt]

\midrule \addlinespace [10pt]

\multicolumn{1}{l|}{\multirow{2}{*}{Anthropogenic}} 
& \multirow{2}{*}{0.91}  & \textbf{43.7} & \multicolumn{1}{c|}{-18.9} & 

  \cellcolor{gray!15} 4 & \cellcolor{gray!15} -0.16** &
\cellcolor{gray!15} \multirow{1}{*}{\textbf{6.5}} & \cellcolor{gray!15} \multirow{1}{*}{5.0}  & \cellcolor{gray!15} \multirow{1}{*}{4.9}  & \cellcolor{gray!15} \multirow{1}{*}{\textbf{13.5}}  \\ 

\multicolumn{1}{l|}{}&  & (39.9) & \multicolumn{1}{c|}{(-0.6)} & 
 1&  -0.15** &
\multirow{1}{*}{4.2} & \multirow{1}{*}{\textbf{5.1}} & \multirow{1}{*}{2.4}  & \multirow{1}{*}{\textbf{13.7}}  \\ \addlinespace[5pt]

\midrule \addlinespace [10pt]

\multicolumn{1}{l|}{\multirow{2}{*}{\makecell{CO$_2$ - ERF (W/m$^\text{2}$) \\ SMCGL}}}
& \multirow{2}{*}{0.91} &  \textbf{43.5} & \multicolumn{1}{c|}{-13.0} & 

\cellcolor{gray!15}  4 & \cellcolor{gray!15} -0.10  & 
\cellcolor{gray!15} \multirow{1}{*}{5.75} & \cellcolor{gray!15} \multirow{1}{*}{\textbf{9.2}}  & \cellcolor{gray!15} \multirow{1}{*}{5.3}  & \cellcolor{gray!15} \multirow{1}{*}{\textbf{15.4}}  \\ 

\multicolumn{1}{l|}{}& &  (39.0) & \multicolumn{1}{c|}{(-0.3)} & 
 1 & -0.11 &
\multirow{1}{*}{2.1} & \multirow{1}{*}{\textbf{3.1}} & \multirow{1}{*}{1.2}  & \multirow{1}{*}{\textbf{8.0}} \\ \addlinespace[5pt]

\midrule \addlinespace [10pt]

\multicolumn{1}{l|}{\multirow{2}{*}{Aerosol}} 
& \multirow{2}{*}{-0.84} &  \textbf{37.9} & \multicolumn{1}{c|}{-45.7} &
 
\cellcolor{gray!15} 4 &\cellcolor{gray!15}  -0.17** & 
\cellcolor{gray!15} \multirow{1}{*}{\textbf{3.9}} & \cellcolor{gray!15} \multirow{1}{*}{2.9}  & \cellcolor{gray!15} \multirow{1}{*}{\textbf{6.1}}  & \cellcolor{gray!15} \multirow{1}{*}{0.0} \\ 

\multicolumn{1}{l|}{}&  & (19.4) & \multicolumn{1}{c|}{(-1.3)} & 
  1 & -0.00 &
\multirow{1}{*}{2.0} & \multirow{1}{*}{\textbf{33.3}} & \multirow{1}{*}{2.0}  & \multirow{1}{*}{\textbf{33.1}}  \\ \addlinespace[5pt]

\midrule \addlinespace [10pt]

\multicolumn{1}{l|}{\multirow{2}{*}{Solar}} 
& \multirow{2}{*}{31.4} &  \textbf{7.0} & \multicolumn{1}{c|}{2.1} & 

\cellcolor{gray!15}  8 & \cellcolor{gray!15} 0.05 &
\cellcolor{gray!15} \multirow{1}{*}{\textbf{4.6}} & \cellcolor{gray!15} \multirow{1}{*}{0.9}  & \cellcolor{gray!15} \multirow{1}{*}{\textbf{3.1}}  & \cellcolor{gray!15} \multirow{1}{*}{1.4}  \\ 

\multicolumn{1}{l|}{}& &  (1.1) & \multicolumn{1}{c|}{(0.6)} & 
 1 & 0.06 &
\multirow{1}{*}{\textbf{6.7}} & \multirow{1}{*}{1.0} & \multirow{1}{*}{\textbf{4.4}}  & \multirow{1}{*}{2.0}  \\ \addlinespace[5pt]

\midrule \addlinespace [10pt]

\multicolumn{1}{l|}{\multirow{2}{*}{Volcanic}} 
& \multirow{2}{*}{0.11} &  \textbf{1.3} & \multicolumn{1}{c|}{-0.3} & 

\cellcolor{gray!15} 4 & \cellcolor{gray!15} 0.18** & 
\cellcolor{gray!15} \multirow{1}{*}{\textbf{10.1}} &\cellcolor{gray!15}  \multirow{1}{*}{0.5}  & \cellcolor{gray!15} \multirow{1}{*}{\textbf{2.7}}  & \cellcolor{gray!15} \multirow{1}{*}{2.4} \\ 

\multicolumn{1}{l|}{}& &  (0.2) & \multicolumn{1}{c|}{(-0.2)} & 
 1 & 0.20***  &
\multirow{1}{*}{\textbf{7.1}} & \multirow{1}{*}{0.3} & \multirow{1}{*}{0.6}  & \multirow{1}{*}{\textbf{3.7}}  \\ \addlinespace[5pt]

\midrule \addlinespace [10pt]

\multicolumn{1}{l|}{\multirow{2}{*}{PDO}} 
& \multirow{2}{*}{0.15} &  \textbf{-2.3} & \multicolumn{1}{c|}{-0.4} & 

\cellcolor{gray!15} 4 & \cellcolor{gray!15} 0.4*** & 
\cellcolor{gray!15} \multirow{1}{*}{\textbf{31.0}} & \cellcolor{gray!15} \multirow{1}{*}{0.8} & \cellcolor{gray!15} \multirow{1}{*}{3.9}  & \cellcolor{gray!15} \cellcolor{gray!15} \multirow{1}{*}{\textbf{13.7}}   \\ 

\multicolumn{1}{l|}{}& &  (-0.3) & \multicolumn{1}{c|}{(-0.3)} & 
1 & 0.38*** & 
\multirow{1}{*}{\textbf{9.5}} & \multirow{1}{*}{0.3} & \multirow{1}{*}{0.7}  & \multirow{1}{*}{\textbf{13.5}} \\ \addlinespace[5pt]

\midrule \addlinespace [10pt]

\multicolumn{1}{l|}{\multirow{2}{*}{\makecell{CO$_2$ (Mt/yr)}}} 
& \multirow{2}{*}{0.89} &  \textbf{42.0} & \multicolumn{1}{c|}{1.0} & 

\cellcolor{gray!15} 2 & \cellcolor{gray!15}  -0.12 & 
\cellcolor{gray!15} \multirow{1}{*}{\textbf{7.9}} & \cellcolor{gray!15} \multirow{1}{*}{1.9}  & \cellcolor{gray!15} \cellcolor{gray!15} \multirow{1}{*}{\textbf{8.8}}  & \cellcolor{gray!15} \multirow{1}{*}{0.3} \\ 

\multicolumn{1}{l|}{}& &  (31.0) & \multicolumn{1}{c|}{(0.00)} & 
 1  &  -0.10  & 
\multirow{1}{*}{\textbf{5.4}} & \multirow{1}{*}{0.0} & \multirow{1}{*}{\textbf{5.4}}  & \multirow{1}{*}{0.7} \\ \addlinespace[5pt]

\midrule \addlinespace [10pt]

\multicolumn{1}{l|}{\multirow{2}{*}{\makecell{CO$_2$ (W/m$^\text{2}$)}}} 
& \multirow{2}{*}{0.91} &  \textbf{43.4} & \multicolumn{1}{c|}{-13.4} & 

\cellcolor{gray!15} 2 & \cellcolor{gray!15} 0.18** & 
\cellcolor{gray!15} \multirow{1}{*}{5.0} & \cellcolor{gray!15} \multirow{1}{*}{\textbf{16.1}}  & \cellcolor{gray!15} \cellcolor{gray!15} \multirow{1}{*}{6.1}  & \cellcolor{gray!15} \multirow{1}{*}{\textbf{6.2}} \\ 

\multicolumn{1}{l|}{}& &  (38.3) & \multicolumn{1}{c|}{(-0.3)} & 
1 & 0.06 & 
\multirow{1}{*}{1.5} & \multirow{1}{*}{\textbf{3.4}} & \multirow{1}{*}{1.0}  & \multirow{1}{*}{\textbf{1.6}} \\ \addlinespace[5pt]

\bottomrule	 \bottomrule
			\end{tabular}
			\end{adjustbox}
		}
	\end{center}
	
	\begin{spacing}{1.0} \footnotesize \noindent Notes: $i$ corresponds to the type of radiative forcing, listed in the left most column. The second column ("Correlation") documents the correlation between GMTA and variable $i$. FEVD values are taken at horizon $h = 15$, which translates into the contribution of variable $i$ in the variance of the forecast error of variable $j$ a decade and a half after the in-sample end date. Numbers in bold underline the highest absolute causal flow among a ($i$,GMTA) pair for a given measure. "*", "**", and "***" means that the null of the residuals cross-correlation of residuals is rejected at the 10\%,5\% and 1\% level respectively. Sample period: 1850-2017.
	\end{spacing}\ \\
\end{table}

\begin{table} [h!]
\renewcommand{\arraystretch}{1.2}
	\caption{Transient Climate Response \\ {\footnotesize Sample Period: 1850-\textbf{2017}}}
	\vspace*{-0.5cm}
	\label{tab:TCR_18502017}
	\begin{center}
		{ \footnotesize
		\begin{adjustbox}{max width=\textwidth}
			\begin{tabular}{l c cc  c cc}
				\toprule \toprule \addlinespace[5pt]
				
 \multicolumn{1}{c}{\multirow{2}{*}{Ordering}} 	& &   \multicolumn{2}{c}{\multirow{1}{*}{Without Trend}} & &  \multicolumn{2}{c}{\multirow{1}{*}{With Trend}}\\	
				\cmidrule(lr){3-4} \cmidrule(lr){6-7} \addlinespace[1pt]
				
	& &    \multicolumn{1}{c}{\multirow{1}{*}{TCR$_{20}$}} &  \multicolumn{1}{c}{\multirow{1}{*}{TCR$_{70}$}} & &   \multicolumn{1}{c}{\multirow{1}{*}{TCR$_{20}$}} &  \multicolumn{1}{c}{\multirow{1}{*}{TCR$_{70}$}} \\			
 \cmidrule(lr){1-1} \cmidrule(lr){3-3} \cmidrule(lr){4-4}  \cmidrule(lr){6-6} \cmidrule(lr){7-7}  \addlinespace[5pt]
				 
 \multicolumn{1}{c}{CO$_2$, GMTA} & & 1.46$^{\circ}$C & 1.94$^{\circ}$C & & 1.76$^{\circ}$C & 2.35$^{\circ}$C \\ 	 \addlinespace[5pt]		
 
  \multicolumn{1}{c}{GMTA, CO$_2$} & & 0.58$^{\circ}$C & 1.79$^{\circ}$C & & 0.97$^{\circ}$C & 2.22$^{\circ}$C \\ 	 \addlinespace[1pt]			

\bottomrule	 \bottomrule
			\end{tabular}
			\end{adjustbox}
		}
	\end{center}
\end{table}

\end{document}